\documentclass[aps,prd]{revtex4}
\usepackage{amsmath}
\usepackage{epsfig}
\newcommand{\be}{\begin{equation}}
\newcommand{\ee}{\end{equation}}
\newcommand{\bea}{\begin{eqnarray}}
\newcommand{\eea}{\end{eqnarray}}

\newcommand{\slashs}[1]{\not{\!#1}}

\def\nnb{\nonumber}
\def\ep{\eta^{\prime}}

\begin{document}

\title{  B Physics and Hadronic Matrix Elements
\footnote{Summary of the Workshop on Advanced Study for B-Physics, Dalian, China}}

\author{Chao-Hsi Chang$^a$, Kuang-Ta Chao$^b$, Xiao-Gang He$^b$\footnote{On leave of absence from
Taiwan University, Taipei}, Chao-Shang Huang$^a$, Tao Huang$^c$,
Zuo-Hong Li$^d$, Cai-Dian L\"u$^c$, Cong-Feng Qiao$^e$, Xing-Hua
Wu$^c$, Yue-Liang Wu$^a$, Zhen-Jun Xiao$^f$,    Ya-Dong Yang$^g$,
  Xian-Qiao Yu$^c$}

 \affiliation{
$^a$   Institute of
     Theoretical Physics, Chinese Academy of sciences, Beijing 100080,
     China\\
 $^b$ Department of Physics, Peking University, Beijing 100871, China\\
 $^c$ Institute of High Energy Physics, Chinese Academy of Sciences,
P.O.Box 918(4), Beijing 100039, China\\
$^d$ Department of Physics, Yantai University, Yantai, 264005,
China\\
$^e$ Department of Physics, Graduate School of the Chinese Academy
of Sciences, Beijing 100039, China\\
$^f$ Department of Physics, Nanjing Normal University, Nanjing,
Jiangsu 210097, P.R.China  \\
$^g$ Physics Department, Henan Normal University, XinXiang, Henan
453007}

\begin{abstract}
In a mini workshop on B Physics which were held at Dalian, China,
a number of lectures was given. Around 40 participants joined the
workshop. In this workshop, people discussed the heavy quark
effective field theory, light cone QCD sum rules and light cone wave
functions. Much discussion was devoted to the currently popular
QCD factorization and perturbative QCD approaches in the hadronic
B decays. The physics of CP violation and physics beyond standard
model were also discussed.
\end{abstract}


\maketitle

\renewcommand{\theequation}%
 {\arabic{equation}}


\section{Outline}

The standard model (SM) has been well tested in the gauge sector,
the unsolved and unclear problems in the SM are mainly concerned
in the Yukawa sector which is strongly related to flavor physics,
such as: origin and mechanism of CP violation, origin of the quark
and lepton masses as well as their mixing. It involves thirteen
parameters whose origin are all unknown. Therefore, precisely
extracting those parameters and testing CP violation mechanism as
well as probing new physics become hot topics in flavor physics.
In fact, flavor physics has already indicated the existence of new
physics. After forty years of discovery for indirect CP violation,
direct CP violation in kaon decays has well been established.
Theoretical predictions \cite{ylw} and experimental measurements
\cite{NA48,KTeV} are now consistent each other. Exclusive
semi-leptonic and inclusive B decays play a crucial role for
extracting two important parameters $V_{cb}$ and $V_{ub}$ in the
CKM matrix elements. Rare B decays and direct CP violations are
also of great importance in determining weak phase angles of the
unitarity triangle and testing the Kobayashi-Maskawa (KM)
mechanism \cite{kobayashi:1973fv} in SM as well as probing new
physics.

 The currently running B-Factories at KEK and SLAC provide more and
 more experimental data on B physics. The future  LHCb, BTeV and
 super-B factory will give much more data. More Precise Experimental Data
and more precision theoretical prediction will encourage precision
test of SM at B sector.

In this workshop (next section), Y.L. Wu talks about how the heavy
quark effective field theory provides a powerful tool for studying
beauty physics, what is the implication of beauty physics. Some
attentions are also paid to the more precise extraction of
$V_{cb}$ and $V_{ub}$ and large direct CP violation in charmless B
decays.

In section \ref{sum}, T. Huang gives a  brief review on the
light-cone wave function(LCWF) in QCD. The definition of the
light-cone wave function in QCD and  distribution amplitude  and
its asymptotic form are reviewed. General properties of the
light-cone wave function and higher helicity components are
discussed.  C.F. Qiao discusses about the derivation of B meson
wave function. X.H. Wu talks about the twist 3 wave function of
pion in QCD sum rules. The $1/m_b$ power suppressed effects are
discussed by Z.H. Li.

Y.D. Yang introduces the QCD factorization approach and discusses
the application in hadronic B decays in section \ref{qcdf}. K.T.
Chao intensively studies the B decays to charmonium final states
in QCD factorization in section \ref{chao}. X.Q. Yu and C.D. Lu
present the formalism and application of perturbative QCD
approach, which is based on $k_T$ factorization in section
\ref{pqcd}. Intensive discussions are induced for the comparison
and underlining theory of these two approaches.

The CKM angle determination is one of the main talks for the B
factories. Z.J. Xiao discusses a number of channels which are
useful for these  angles $\alpha$, $\beta$ and $\gamma$ in section
\ref{ckm}. In section \ref{bs},  C.H. Chang discuss the exact
solution of Bethe-Salpeter  (BS) equation.

 The new physics picture is always a hot topic. X.G.
He and C.S. Huang give the SUSY contributions in $B\to K^* \gamma$
decay and $B\to \phi K_S$ and $B\to \eta' K_S$ decays in section
\ref{new}. These decays are more sensitive to new physics than
some others. Z.J. Xiao makes a systematic study for the new
physics contributions to the charmless two-body hadronic decays of
B and $B_s$ meson induced by the neutral- and charged-Higgs
penguin diagrams in model III: the third type of the
two-Higgs-doublet models and technicolor models.

\section{LCQCD/HQEFT From Full QCD }
    \centerline{by Y.L. Wu} \hspace{.8cm}

 Heavy quark effective field theory  of QCD which was
first explored in \cite{W0} and detailed developed recently in a
serious papers \cite{W1,W2,W3,W4,W5,W6,W7,W8,W9,W10,W11,W12}
provides a premising and systematic tool in correctly evaluating
the hadronic matrix elements of heavy quarks and extracting the
CKM matrix elements $V_{cb}$ and $V_{ub}$ from B decays via heavy
quark expansion (HQE). HQEFT is a theoretical framework derived
directly from QCD but explicitly displays the heavy quark symmetry
in the infinite mass limit $m_Q\rightarrow \infty$ and symmetry
breaking corrections for finite mass case in the real world. In
fact, HQEFT has been shown to be as a large component QCD
\cite{W0,W10}. At the leading order, it coincides with the usual
heavy quark effective theory(HQET) \cite{HQET} which is
constructed based on the heavy quark symmetry (HQS) \cite{HQS} in
the infinite mass limit \cite{HQL}. The differences between HQEFT
of QCD and HQET arise from the sub-leading terms in the $1/m_Q$
expansion. This is because in the construction of HQET the
particle and antiparticle components were separately treated based
on the assumption that the particle number and antiparticle number
are conserved in the effective Lagrangian \cite{MR}. However, such
an assumption is only valid in the infinite mass limit. Note that
the particle number and antiparticle number are always conserved
in the transition matrix, which is independent of heavy quark
limit. Therefore the particle-antiparticle coupling terms which
correspond to the pair creation and annihilation terms in full QCD
were inappropriately dropped away in the HQET though they are
suppressed by $1/m_Q$ but only vanish in infinite mass limit. In
HQEFT of QCD \cite{W0,W10}, all contributions of the field
components, large and small, `particle' and `antiparticle', have
carefully been treated in the effective Lagrangian, so that the
resulting effective Lagrangian form the basis for a ``complete"
effective field theory of heavy quarks. It is seen that the
particle-antiparticle coupling terms are suppressed by $1/m_Q$ and
they are decoupled in the infinite mass limits. Their physics
effects at sub-leading order have been found to be significant in
some cases \cite{W1,W2,W3,W4,W5,W6,W7,W8,W9}. Therefore, to
consider the finite quark mass correction precisely, it is
necessary to include the contributions from the components of the
anti-quark fields. Note that the Wilson coefficient functions
describe the perturbative effects in full QCD, HQEFT of QCD treats
the non-perturbation effects below the energy scale $m_b$. So that
the anti-quark effects in HQEFT of QCD cannot be attributed to the
Wilson coefficient functions. As the anti-quark effects appear
 from the sub-leading order of $1/m_Q$ expansion, it should not
surprise that at the leading order the resulting anomalous
dimensions from both full QCD and HQEFT of QCD are all the same.

 A complete description for the theoretical framework of
LCQCD/HQEFT is going to be presented in a longer paper with the
following contents:

\noindent
1.\ Introduction \\
2.\ Effective Lagrangian of LCQCD/HQEFT from QCD \\
3.\ Lorentz Invariance of LCQCD/HQEFT \\
4.\ Quantization of LCQCD/HQEFT \\
  $\hspace{1.0cm}$ 4.1\ Quantum Generators of Poincare Group \\
\  $\quad$  4.2\ Anti-commutations and Super selection Rule \\
\  $\quad$  4.3\ Hilbert and Fock Space of LCQCD/HQEFT \\
\  $\quad$  4.4\ Propagator in LCQCD/HQEFT\\
\  $\quad$  4.5\ Discrete Symmetries in LCQCD/HQEFT \\
5.\ Basic Framework of LCQCD/HQEFT \\
\  $\quad$  5.1\ Feynman Rules in LCQCD/HQEFT \\
\  $\quad$  5.2\ Finite Mass Corrections \\
\  $\quad$  5.3\ Trivialization of Gluon Couplings and Decouple
Theorem \\
\  $\quad$  5.4\ Renormalization in LCQCD/HQEFT and Wilson Loop \\
\  $\quad$  5.5\ Current Operators in LCQCD/HQEFT
\\
6.\ New Symmetries in LCQCD/HQEFT \\
7. Demonstration for the Trace Formula of Transition Matrix
Elements and Universal Isgur-Wise Function \\
 8. Simple applications

Hereafter is a summary of Important Effects in LCQCD/HQEFT:
 \begin{itemize}
  \item Automatically reparameterization invariance
  $v\to v'$ with $v^2 = v^{'2} = 1$.
  \item Automatically with vanishing $1/m_Q$ corrections
  at Zero-Recoil (both for $B\to D^{\ast} l \nu_l $ $\&$
  $B\to D l \nu_l$ )
  \item Largely reduce the numbers of form factors
   at higher Order of $1/m_Q$ and more relations are
   obtained
  \item Nicely keep scaling law of decay constants

  $f_{H} = \frac{F}{\sqrt{M_H} } [ 1 + O(\frac{1}{m_Q}
 )] $
 $\quad$ ($\sim 10\%$ \ for b ,\ $\sim 30\%$ for c )

  The explicit forms and results were found to be as follows
 \begin{eqnarray*}
 \label{decayrelationwithg}
 & & f_M=\frac{F}{\sqrt{m_M}} \{ 1+\frac{1}{m_Q}(g_1+2d_M g_2)\}
 \\
 & & \frac{f_V m^{1/2}_V}{f_P m^{1/2}_P}=1-\frac{8}{m_Q} g_2 \\
 & & F = 0.38\pm 0.06 \; \mbox{GeV}^{3/2}; \quad g_1 = 0.46 \pm 0.12 \;
 \mbox{GeV} ; \quad g_2 = -0.06\pm 0.02 \; \mbox{GeV}
 \end{eqnarray*}
 The numerical results are
 \begin{eqnarray*}
  \label{eq:decval3QCD2loop}
   &f_B(m_b)=0.196 \pm 0.044  \; \mbox{GeV}, \hspace{2cm}
   &f_{B^{\ast}}(m_b)=0.206 \pm 0.039 \;  \mbox{GeV},\nonumber\\
   &f_D(m_c)=0.298 \pm 0.109  \; \mbox{GeV}, \hspace{2cm}
   &f_{D^{\ast}}(m_c)=0.354 \pm 0.090 \;  \mbox{GeV}.
 \end{eqnarray*}
  \item Reliably understand the lifetime differences of
 bottom hadrons $B_d$, $B_s$, $\Lambda_b$.
 \begin{eqnarray*}
& & \frac{\tau(\Lambda_b)}{\tau(B^0)}=0.79\pm 0.01 \\
& & \frac{\tau (B^-)}{\tau(B_d)} = 1.08 \pm 0.05\, ,\qquad \frac{\tau (B_s)}{\tau(B_d)} = 0.96 \pm 0.06\, \\
& & B_{SL}(B^0)= (10.51 \pm 0.37), \qquad B_{SL}(\Lambda_b)=(11.13 \pm 0.26 ) \% \\
& & n_c(B^0)=1.20 \pm 0.03, \qquad B_{\tau}(B^0)= (2.6 \pm 0.1) \%
\end{eqnarray*}
  \item Consistently extraction of $V_{cb}$ and $V_{ub}$
\begin{eqnarray*}
& & |V_{cb}|=0.0395 \pm 0.0011_{\mbox{exp}}\pm 0.0019_{\mbox{th}}
\qquad \mbox{from} \quad  B\to D^*l\nu ,
\quad O(1/m^2_Q) \\
& & |V_{cb}|=0.0434 \pm 0.0041_{\mbox{exp}}\pm 0.0020_{\mbox{th}}
\qquad \mbox{from} \quad B\to Dl\nu ,
\quad O(1/m^2_Q) \\
& &|V_{cb}|=0.0394 \pm 0.0010_{\mbox{exp}}\pm 0.0014_{\mbox{th}}
\qquad \mbox{from} \quad B\to X_c l\nu, \quad O(1/m^2_Q) \\
& & |V_{cb}|=0.0402 \pm 0.0014_{\mbox{exp}} \pm 0.0017_{\mbox{th}}
\qquad (Average) \\
& &  |V_{ub}|_{LO}=(3.4\pm 0.5_{\mbox{exp}} \pm 0.5_{\mbox{th}})
   \times 10^{-3}  \qquad \mbox{from} \quad B\to \pi l\nu  ,\\
& & |V_{ub}|_{LO}=(3.7\pm 0.6_{\mbox{exp}} \pm 0.7_{\mbox{th}}
)\times 10^{-3} \qquad \mbox{from} \quad B\to \rho l\nu ,\\
& & |V_{ub}|_{NLO}= (3.2 \pm 0.5_{\mbox{exp}} \pm
0.2_{\mbox{th}})\times 10^{-3}
\qquad \mbox{from} \quad  B\to \pi l\nu, \\
& &  |V_{ub}|=(3.48 \pm 0.62_{\mbox{exp}}\pm
0.11_{\mbox{th}})\times 10^{-3}  \qquad \mbox{from} \quad  B\to
X_u l\nu  \quad O(1/m^2_Q)
\end{eqnarray*}

\end{itemize}

\section{Light-cone Sum Rules and Light-cone Wave Function in $QCD$}
\label{sum}  \centerline{by T. Huang, Z.H. Li, C.F. Qiao and X.H.
Wu} \hspace{.8cm}

 A brief review on the light-cone
wave function  in QCD is presented.It consists of the following
nine sections: 1) the definition of the light-cone wave function
in QCD; 2) distribution amplitude(DA) and its asymptotic form; 3)
general properties of the light-cone wave function; 4) Melosh
transformation and higher helicity components; 5) DA moments in
the QCD sum rules; 6) model approach to LCWF; 7) Conformal
expansion; 8) Twist-3 distribution amplitude of the pion; 9)
perspective on the light-cone wave function.

As well known,the light cone formalism provides a convenient
framework for the relativistic description of hadrons in terms of
quark and gluon degrees of freedom, and the application of
perturbative QCD (PQCD) to exclusive processes has mainly been
developed in this formalism \cite{lb}.In the PQCD theory, the
hadronic distribution amplitudes and structures which enter
exclusive and inclusive processes via the factorization theorems
at high momentum transfer can be determined by the hadronic wave
functions, and they are the underlying links between hadronic
phenomena in QCD at large distances (nonperturbative) and small
distance (perturbative). A light-cone (LC) wave function is
defined at the LC time $\tau\equiv x^{+}=x^{0}+x^{3}$ in physical
light cone gauge $A^+=A^0+A^3=0$, which is conjugate to the LC
Hamiltonian $H_{LC}\equiv P^{-}=P^{0}-P^{3}$ \cite{bhl}.A
stationary solution $|\Psi(\tau)\rangle$ has a LC Fock state
expansion and the LC wave functions, $\Psi_n(x_i,{\bf k}_{\perp
i},\lambda_i)$, are the amplitudes to find $n$ partons (quarks,
anti-quarks and gluons) with momenta $k_i$ in a pion of momentum
$P$. Hence it is important to develop methods which specify the
light-cone wave functions of the hadron.

The distribution amplitude $\Phi(x,Q)$ for the leading twist
satisfies a QCD evolution equation \cite{lb}.As $Q^{2}$ goes to
infinity,$\Phi(x,Q)$ is dominated mainly by the first term in the
expansion,i.e for the twist-2 DA of the meson,we have
$\Phi_{M}(x,Q)\rightarrow a_{M}^{(0)}x(1-x)$.However the solution
of the evolution equation can be determined completely only as if
the initial condition of $\Phi(x,Q_{0})$ is given.

 From the relation between the wave functions and measurable
quantities we can get the general properties of the hadronic wave
functions \cite{bhl}. For example,in the pion case two important
constraints on the valence state wave function
$\Psi_{q\bar{q}}(x,{\bf k}_{\perp})$ (for the
$\lambda_1+\lambda_2=0$ components) have been derived from
$\pi\to\mu\nu$ and $\pi^0\rightarrow \gamma\gamma$ decay
amplitude. The quark transverse momentum of the valence state in
the pion $\langle \mathbf{k}_{\perp}^2\rangle_{q\bar{q}}$ should
be larger than $\langle\mathbf{k}_{\perp}^2 \rangle_{\pi}$.
$P_{q\bar{q}}$ is the probability of finding the $q\bar{q}$ Fock
state in the pion, and it should be less than unity. the charged
pion radius for the lowest valence quark state $\langle
r_{\pi^+}^2 \rangle^{q\bar{q}}$ should be lower than the
experimental value for $\langle r_{\pi^+}^2 \rangle$, where the
latter one should contain all the contributions of the higher Fock
stats.

In order to obtain the LC spin wave function of the pion, we
transform the ordinary instant-form SU(6) quark model wave
function into a LC one. The instant-form spin states
$|J,s\rangle_T$ and the LC-form spin states $|J,\lambda\rangle_F$
are related by $|J,\lambda\rangle_F=\sum_s U_{s\lambda}^J
|J,s\rangle_T$ . This rotation is called the Melosh rotation for
spin-1/2 particles. Applying the Melosh rotation  we can obtain
the spin wave function of the pion in the infinite-momentum frame
 \cite{mh}. There are two higher helicity components
$(\lambda_1+\lambda_2 = \pm 1)$ in the expression of the LC spin
wave function besides the ordinary helicity components
$(\lambda_1+\lambda_2 =0)$. These higher helicity components will
make contributions to the exclusive processes although they have
the power suppression behavior.

Applying the QCD sum rules to the distribution amplitude,we can
get the first three moments of the initial DA,$\Phi(x,Q_{0})$
\cite{sumrule}. From these moments we can not determine the
initial $\Phi(x,Q_{0})$ and only know that the initial
$\Phi(x,Q_{0})$ is different from the asymptotic form.

For the space wave function,one can take the Brodsky-Huang-Lepage
(BHL) prescription which is obtained by connecting the equal-time
wave function in the rest frame and the wave function in the
infinite momentum frame \cite{bhl}. We obtain the LC space wave
function of the pion, $\varphi_{\mathrm{BHL}}=A \exp[-\frac{{\bf
k}_{\perp}^2
  +m^2}{8{\beta}^2x(1-x)}]$  \cite{bhl}. This form provides a more
reasonable $k_{T}$ dependence in the space wave function and is
consistent with the experimental data \cite{hs}.

Conformal expansion allows one to separate transverse and
longitudinal variables in the distribution amplitude.one can
expand DA in terms of Gegenbauer polynomials $C_{n}^{1/2}$ and
$C_{n}^{3/2}$ according to its conformal spin  \cite{bf}. This
symmetry helps us to make a model for the light-cone wave function
of the hadron.

Recently, we do not use the equation of motion for the quarks
inside the pion and apply the QCD sum rules to calculate the pion
twist-3 DA $\Phi_{p}(\xi)$ and get its approximate expression and
its expansion coefficients can be determined by its three moments
approximately  \cite{hwz}. Based on the moment calculation we
suggest a model for the twist-3 wave function with $k_{T}$
dependence  \cite{hw}.

In Summary, at present although we can not solve light-cone wave
function of the hadron completely due to its non-perturbative
dynamics, one has got much progress by studying its general
properties,symmetry, its moments from non-perturbative calculation
and phenomenological model. These studies will contribute to a
better understanding on the light-cone wave functions and approach
their realistic solutions.

\subsection{Twist-$3$ Distribute Amplitude of the Pion in QCD Sum
Rules}

We apply the background field method to calculate the moments of
the pion two-particle twist-3 distribution amplitude (DA)
$\phi_p(\xi)$ in QCD sum rules. The QCD sum rules were used to
study the leading-twist distribution amplitude of the pion at the
first time in  \cite{sumrule}. The pion twist-3 distribution
amplitudes have been studied in Ref.  \cite{Braun2,Ball} in the
chiral limit, based on the techniques of nonlocal product
expansion and conformal expansion and including corrections in the
meson-mass. However they employ the equations of motion of
on-shell quarks in the meson to get two relations among
two-particles twist-3 distribution amplitudes of the pion,
$\phi_p(\xi)$ and $\phi_\sigma(\xi)$, and three-particle twist-3
distribution amplitude $\phi_{3\pi}(\alpha_i)$ of the pion. The
question is whether one can use the equation of motion of the
quark inside the meson since the quarks are not on-shell. In our
paper \cite{hwz}, we do not apply the quark equation of motion and
calculate the moments of the twist-3 distribution amplitudes
$\phi_p(\xi)$ directly in QCD sum rules approach.

For calculating the moments of $\phi_p(\xi)$ , we take as usual
the two-point correlation functions:
\begin{eqnarray*}
i\int d^4 x e^{iq\cdot x} \langle 0\mid T \left \{ j_5^{(2n)}(x,z)
j_5^{(0)\dag}(0) \right \} \mid 0 \rangle \equiv (z\cdot
q)^{2n}I_p^{2n,0}(q^2),
\end{eqnarray*}
 where the currents are defined as
\begin{eqnarray*}
& & j_5^{(2n)}(x,z)=\bar{d}(x)\gamma_5(iz\cdot
\overleftrightarrow{D})^{2n} u(x)
\end{eqnarray*}
with$\overleftrightarrow{D}=\overleftrightarrow{\partial_\mu}-i2gA_\mu^a
T^a$. To get the sum rules for moments, we employ dispersion
relation for $I_p^{2n,0}(Q^2)$ and isolate the pole term of the
lowest pseudoscalar pion. With condensates up to dimension-6 and
the perturbative contribution part to the lowest order, we obtain
the sum rule for moments of $ \phi_p(\xi) $:
\begin{eqnarray*}
&&\langle \xi_p^{2n} \rangle \cdot \langle \xi^0_p \rangle
=\frac{M^4}{(m_0^p)^2 f^2_{\pi}} e^{m_{\pi}^2/M^2}
 \left[ \frac{3}{8\pi^2}\frac{1}{2n+1} \left(
    1-(1+\frac{s_{\pi}}{M^2})e^{-\frac{s_{\pi}}{M^2}}\right) \right.
\nonumber \\
& &-(2n-1)\frac{\bar m \langle \bar{\psi}\psi \rangle}{M^4}
 +\frac{2n+3}{24} \frac{\langle
    \frac{ \alpha_s}{\pi}G^2 \rangle}{M^4} \left.  -\frac{16
    \pi}{81}\left( 21 + 8n(n+1) \right) \frac{\langle
    \sqrt{\alpha_s}\bar{\psi}\psi \rangle^2}{M^6} \right]
\end{eqnarray*}
with $ \bar m=(m_u+m_d)/2 $, and $ M $ is a Borel parameter. For
the condensate parameters, we take as usual. And the definition of
the moments are given by
\begin{eqnarray*}
\langle \xi_p^{2n}
\rangle=\frac{1}{2}\int_{-1}^{1}\xi^{2n}\phi_p(\xi)d\xi .
\end{eqnarray*}

The key point to avoid the concept of on-shell equations of motion
is that the introduced $m_0^p$ appears in the sum rule of moments.
And they can be obtained through requiring normalization of the
zeroth moments($\langle \xi^0_p \rangle =1$). The advantage is
that our results do not rely on the validness of the equation of
motion and at the same time provide a parameter which can be
compared with the corresponding one used in phenomenological
analysis. With these points kept in mind, one can show that the
results for the first three moments are($M^2=1.5-2~{\rm GeV}^2$):
\begin{eqnarray}
\begin{array}{lllllll}
s_\pi({\rm GeV}^2) & & m_0^p({\rm GeV}) & & \langle \xi^2_p \rangle & & \langle \xi^4_p \rangle \\
1.7 & & 1.24-1.36 & & 0.340-0.356 & & 0.167-0.210 , \\
1.6 & & 1.19-1.30 & & 0.340-0.359 & & 0.164-0.211 , \\
1.5 & & 1.14-1.24 & & 0.341-0.361 & & 0.160-0.212 .
\end{array}
\nonumber
\end{eqnarray}
It can be seen that, the parameter $m_0^p$ is smaller than
$m_\pi^2/(m_u+m_d)\simeq 1.78-3.92~{\rm GeV}$ which was used in,
e.g.,  \cite{Braun2} when the equation of motion of on-shell
quarks are employed. We take the first three moments into
consideration in our analysis. Expanding twist-3 distribution
amplitudes $\phi_p(\xi)$ of the pion in Gegenbauer polynomials to
first three terms, one can obtain its approximate form,
\begin{equation*}
 \phi_p(\xi)=1 + 0.137 ~ C^{1/2}_2(\xi)-0.721 ~ C^{1/2}_4(\xi).
\end{equation*}

\subsection{B-Meson Wave Functions}

The development of the heavy quark effective theory(HQET) has led
to much progress in the theoretical understanding of the
properties of hadrons. With the progress in both theory and
experiment, nowadays, $B$ physics becomes one of the most active
research areas in high energy physics. Many $B$ meson exclusive
decay processes turn out to be calculable systematically in the
frameworks of newly developed factorization formalisms, the
so-called PQCD approach  \cite{KLS} and QCD Factorization approach
\cite{beneke}. In all of these calculations based on the
factorization approaches, the light-cone distribution amplitudes
of the participating mesons, which express the nonperturbative
long-distance contributions in the factorized amplitudes, play an
important role in making reliable predictions.

By definition, the light-cone distribution amplitudes are given by
the light-cone  wave functions at zero transverse separation of
the constituents. Following  \cite{grozin} it is
\bea \langle 0 | \bar{q}(z) \Gamma h_{v}(0) |\bar{B}(p) \rangle =
- \frac{i f_{B} M}{2}
{\rm Tr}
 \left[ \gamma_{5}\Gamma \frac{1 + \not\!{v}}{2}
 \left\{ \tilde{\phi}_{+}(t) - \not\!{z} \frac{\tilde{\phi}_{+}(t)
 -\tilde{\phi}_{-}(t)}{2t}\right\} \right].
\label{phi1} \eea
Eq. (\ref{phi1}) is the most general parametrization of the
two-particle light-cone matrix element compatible with Lorentz
invariance and the heavy-quark limit. The three-particle matrix
element can be generally expressed as \cite{qiao}:
\bea &&\hspace{-0.7cm}\lefteqn {\langle 0 | \bar{q} (z) \, g
G_{\mu\nu} (uz)\, z^{\nu}\, \Gamma \, h_{v} (0) | \bar{B}(p)
\rangle = \frac{1}{2}\, f_B M \, {\rm Tr}\, \left[ \,
\gamma_5\,\Gamma \, \frac{1 + \slashs{v}}{2}\, \right.}
\nonumber \\
&&\hspace{-0.7cm}\biggl\{ ( v_{\mu}\slashs{z}
         - t \, \gamma_{\mu} )\  \left( \tilde{\Psi}_A (t,u)
   - \tilde{\Psi}_V (t,u) \right)
- i \, \sigma_{\mu\nu} z^{\nu}\, \tilde{\Psi}_V (t,u)
       -  z_{\mu} \, \tilde{X}_A (t,u)
       + \frac{z_{\mu}}{t} \, \slashs{z} \,\tilde{Y}_A \,(t\,,\,u)
\left. \biggr\} \, \right]\ .~~~~~~\eea
By virtue of the equations of motion, one can get a set of
relations between distribution amplitudes. In the approximation
that the three-particle amplitudes are set to be zero, the
Wandzura-Wilczek approximation, the system of the obtained
differential equations is simplified, and the solutions for
$\tilde{\phi}_{\pm}$ can be obtained. In the momentum space, the
analytic solution can be expressed explicitly as  \cite{qiao}
\bea \label{eq:12} \psi_{+}(\omega, \mbox{\boldmath $k$}_{T}) &=&
\frac{\omega} {2 \pi \bar{\Lambda}^2} \theta(\omega)\theta(2
\bar{\Lambda} - \omega) \delta \left(\mbox{\boldmath $k$}_{T}^{2}
- \omega (2 \bar{\Lambda} - \omega) \right) \eea \bea
\label{eq:13} \psi_{-}(\omega, \mbox{\boldmath $k$}_{T}) &=&
\frac{2 \bar{\Lambda} - \omega}{2 \pi \bar{\Lambda}^2}
\theta(\omega)\theta(2 \bar{\Lambda} - \omega) \delta
\left(\mbox{\boldmath $k$}_{T}^{2} - \omega (2 \bar{\Lambda} -
\omega) \right) \eea
The results (\ref{eq:12}) and (\ref{eq:13}) give exact description
of the valence Fock components of the $B$ meson wave functions in
the heavy-quark limit, and represent their transverse momentum
dependence explicitly. These results show that the dynamics within
the two-particle Fock states is determined solely in terms of a
single nonperturbative parameter $\bar{\Lambda}$.

It should be noted that the Wandzura-Wilczek approximation is not
equivalent to the free field approximation. The leading
Fock-states, which correspond to the twist-2 contribution in the
case of light meson wave functions, carries the effect from the
QCD interaction.

There has been indication that, in the heavy-light quark systems,
the higher Fock states could play important roles even in the
leading twist level  \cite{qiao}. This would suggest that the
shape of the wave functions as function of momenta and their
quantitative role in the phenomenological applications would be
modified when including the higher Fock states. However, the
direct applications of our wave functions to phenomenological
processes indicate that the Wandzura-Wilczek approximation does
not induce obvious conflicts with experimental data  \cite{cdl}

In contrast to many model assumptions, which show strong damping
at large $|\mbox{\boldmath $z$}_{T}|$ as $\sim \exp\left(-
K^2\mbox{\boldmath $z$}_{T}^{2}/2\right)$, our wave functions
(\ref{eq:12}) and (\ref{eq:13}) have slow-damping with oscillatory
behavior as $\psi_{\pm}(\omega, -\mbox{\boldmath $z$}_{T}^{2})
\sim \cos(|\mbox{\boldmath
$z$}_{T}|\sqrt{\omega(2\bar{\Lambda}-\omega)} -\pi/4)
/\sqrt{|\mbox{\boldmath $z$}_{T}|}$. On the other hand,
heavy-quark symmetry guarantees that the solution in our work
provides complete description of the light-cone valence Fock wave
functions for the $B^{*}$ mesons and also for the $D$, $D^{*}$
mesons in the heavy-quark limit. Moreover, the investigations on
beyond Wandzura-Wilczek limit are in progress.

\subsection{$1/m_b$ power-suppressed effects in two-body B
decays}

A careful investigation of $1/m_b$ power suppressed effects is
crucial for a good understanding of two-body hadronic decays of B
mesons. It is instructive to systematically investigate the
$1/m_b$ suppressed effects from either soft or hard gluon
exchanges, in addition to annihilation topology. Focusing on
penguin-dominated $B\to K\pi$ and color-suppressed
$\bar{B}^0\longrightarrow D^{0}\pi^{0}$ decays, we estimate such
power-suppressed corrections which appear possibly in the
framework of QCD light-cone sum rules (LCSR).

In the former case, the soft exchanges could be induced by the
emission topology or by the chromo-magnetic diploe operator
$O_{8g}$ and penguin topology, while the hard gluon exchanges may
occur through the $O_{8g}$ operator or penguin contractions. The
behaviors of such contributions in the heavy quark limit $m_b\to
\infty$ comply with the following power counting: (1) The soft
effects due to the emission topology are of ${\cal O}(1/m_b)$. (2)
The penguin diagrams make no contributions up to $1/m_b^2$ order.
(3) The $O_{8g}$ operator supplies a hard effect of ${\cal
O}(1/m_b)$ and a soft effect of ${\cal O}(1/m_b^2)$. Despite being
formally suppressed by ${\cal O}(1/m_b^2)$ compared with the
leading-order factorizable amplitudes, the soft correction from
$O_{8g}$ is free of $\alpha_s$ suppression and numerically
comparable with the ${\cal O}(\alpha_s)$ hard part. (4) The part
with quark condensate has a chiral enhancement factor
$r_{\chi}^{\pi}$ via the PCAC relation, suffering formally from
${\cal O}(\alpha_s)$ and ${\cal O}(1/m_b)$ double suppression but
numerically turning out to be a large effect. However, it has a
counterpart in the $QCD$ factorization and thus is not included in
our calculation to avoid a double counting. For the weak phase
$\gamma (=\textmd{Im}V_{ub}^*)$ ranging from $40^{\circ}$ to
$80^{\circ}$, we find that because of the annihilation and soft
and hard exchanges a numerical increase of $(20-30)\%$ is expected
in the branching ratios up to ${\cal O}(1/m_b^2)$ corrections; the
resultant soft and hard corrections are less important than the
annihilation contributions and amount only to a level of $10\%$.
Possible sources of uncertainties are discussed in some details.

For the case of $\bar{B}^0\longrightarrow D^{0}\pi^{0}$, naive
factorization assumption breaks down and non-factorizable
contributions are leading. Then the $1/m_b$ power-suppressed soft
exchange between the emitted heavy-light quark pair and $B\pi$
system is expected to be much more important than in the former
case. The resulting correction to the decay amplitude is found to
be comparable numerically with the corresponding factorizable
piece, estimated at about $(50-110)\%$ of the latter, and renders
the relevant parameter $a_2$ receive a positive number correction.
This indicates that such power-suppressed effect would be crucial
to our phenomenological understanding of the
$\bar{B}^0\longrightarrow D^{0}\pi^{0}$ decay.

\section{ QCD Improved Factorization Approach for B Decays}
\centerline{by Y.D. Yang}\hspace{.8cm} \label{qcdf}

It is of great interest and importance to investigate the decays
of B mesons to charmless final states to study the weak
interactions and CP violation. In past years, we have witnessed
many experimental and theoretical progresses in the study of B
physics with the observations of many B charmless decay processes
and improvements of the theory of B decays.

Most of the theoretical studies of B decays to pseudoscalar and
vector final states are based on the popular $Naive$
$Factorization$ approach \cite{BSW}. As it was pointed out years
ago in Ref.\cite{cz}, the dominant contribution in B decays comes
from the so-called $Feynman$ $mechanism$, where the energetic
quark created in the weak decay picks up the soft spectator softly
and carries nearly all of the final-state meson's momentum. It is
also shown that Pion form factor in QCD at intermediate energy
scale is dominated by Feynman mechanism \cite{isg, rad, stef}.
>From this point, we can understand why the naive factorization
approach have worked well for B and D decays, and the many
existing predictions for B decays based on naive factorization and
spectator ansatz do have taken in the dominant physics effects
although there are shortcomings. However, with the many new data
available from CLEO and an abundance of data to arrive within few
years from the B factories BaBar and Belle, it is demanded highly
to go beyond the naive factorization approach.

Recently, Beneke $et.\, al.,$ have formed an interesting QCD
factorization formula for B exclusive non-leptonic decays
\cite{BBNS, beneke}. The factorization formula incorporates
elements of the naive factorization approach(as leading
contribution) and the hard-scattering approach(as sub-leading
corrections), which allows us to calculate systematically
radiative(sub-leading non-factorizable)
 corrections  to naive factorization for B exclusive non-leptonic decays.
An important product of the formula is that the strong final-state
interaction phases are calculable from the first principle which
arise from the hard-scattering kernel and hence process dependent.
The strong phases are very important for studying CP violation in
B decays. Detail proofs and arguments could be found in
\cite{beneke}.

The amplitude of $B$ decays to two light mesons, say $M_1$ and
$M_2$, is obtained through the hadronic matrix element $\langle
M_{1}(p_{1}) M_{2}(p_{2}){\mid}{\cal O}_i {\mid}B(p)\rangle$, here
$M_{1}$ denotes the final meson that picks up the light spectator
quark in the $B$ meson, and $M_{2}$ is the another meson which is
composed of the quarks produced from the weak decay point of $b$
quark. Since the quark pair, forming $M_2$, is ejected from the
decay point of $b$ quark carrying the large energy of order of
$m_b$, soft gluons with the momentum of order of $\Lambda_{QCD}$
decouple from it at leading order of $\Lambda_{QCD}/m_b$ in the
heavy quark limit. As a consequence any interaction between the
quarks of $M_2$ and the quarks out of $M_2$ is hard at leading
power in the heavy quark expansion. On the other hand, the light
spectator quark carries the momentum of the order of
$\Lambda_{QCD}$, and is softly transferred into $M_1$ unless it
undergoes a hard interaction. Any soft interaction between the
spectator quark and other constituents in $B$ and $M_1$ can be put
into the transition form factor of $B\to M_1$. The
non-factorizable contribution to $B\to M_1 M_2$ can be calculated
through the diagrams in Fig.1.

\begin{figure}
 \begin{center}
\includegraphics[width=11cm]{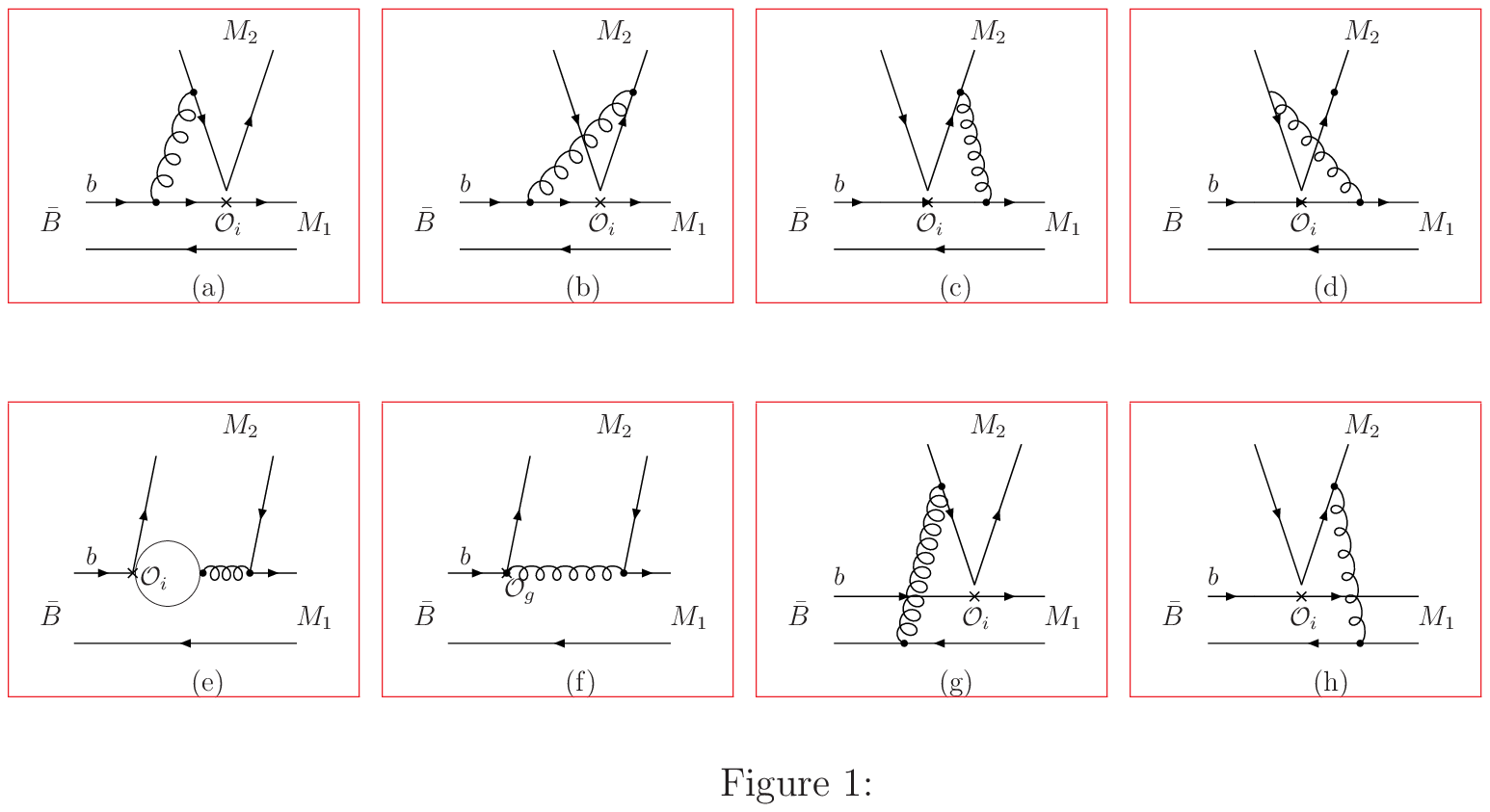}
\end{center}
\caption{Feynman diagrams contributed to non-factorizable contributions.} \label{f1}
\end{figure}

It is worth to note that the shortcomings in the ``generalized
factorization" are resolved in this framework. Non-factorizable
effects are calculated in a rigorous way here instead of being
parameterized by effective color number. Since the hard scattering
kernels are convoluted with the  light cone DAs of the mesons,
gluon virtuality $k^2 ={\bar x}m_b^2$ in the penguin diagram
Fig.1.e has well defined meaning and leaves no ambiguity as to the
value of $k^2$, which has usually been treated as a free
phenomenological parameter in the estimations of the strong phase
generated though the BSS mechanism \cite{BSS}. So that CP
asymmetries could be predicted soundly.

\section{$B$ meson exclusive decays to charmonium in QCD factorization}
\centerline{by K.T. Chao}\hspace{.8cm} \label{chao}

 Exclusive
$B$-meson decays into charmonium  are interesting, since these
decays e.g. $B\to\!J/\psi K$ are regarded as the golden channels
for CP violation studies, and these decays provide useful
information towards the understanding of perturbative and
nonperturbative QCD. These decays are color-suppressed decays, and
involve two heavy quark energy scales, $m_b$ and $m_c$, therefore
are more subtle in theoretical studies. While experimentally CLEO,
BaBar and Belle Collaborations have provided many measurements on
the $B$ meson exclusive decays to charmonium \cite{pdg} such as
$J/\psi, \psi', \eta_c, \eta_c', \chi_{c0}$, and $ \chi_{c1}$,
theoretical studies for those decays are still limited.

In the QCD factorization approach \cite{BBNS,beneke}, it is argued
that because the size of charmonium is small $(\sim
\!\!1/{\alpha_s {m_\psi}})$ and its overlap with the $(B, K)$
system is negligible in the heavy charm quark limit, QCD
factorization method can be used for $B \to\!J/\psi K$ and other
charmonium states, where charmonium is described by the
color-singlet $c\bar c$ pair. Indeed, explicit calculations for $B
\to\!J/\psi K$ within the QCD factorization approach \cite{ch}
show that the nonfactorizable vertex contribution is infrared safe
and the spectator contribution is perturbatively calculable at
twist-2 order, but the theoretical branching ratio is much smaller
than the experimental data. Aside from the  $B \to\!J/\psi K$
decay \cite{ch}, here we present some additional results for $B$
exclusive decays to other S-wave, P-wave, and D-wave charmonium
states \cite{song1, song2, song3, gao} in QCD factorization.
Problem of infrared divergences in $B$ decays to P-wave charmonium
($B \to\! \chi_{c0} K$, $B \to\! \chi_{c2} K$, and $B \to\! h_c
K$) and problem of S-D mixing in $B$ decay to D-wave charmonium
($B \to\! \psi(3770) K$)are emphasized.

\subsection{$B \to\! \eta_c K$ decay} Studies of $B$ meson decays
into pseudoscalar charmonium states $B \to\! \eta_c K$ and  $B
\to\! \eta_c(2S) K$ \cite{song1} in QCD factorization show that
the nonfactorizable corrections to naive factorization are
infrared safe at leading-twist order. The spectator interactions
arising from the kaon twist-3 effects are formally
power-suppressed but chirally and logarithmically enhanced. An
important improvement by including the $\mathcal{O}(\alpha_s)$
corrections is the large cancellation of the renormalization scale
$\mu$ dependence of the decay amplitude. However, the calculated
decay rates are smaller than experimental data by a factor of
8-10. On the other hand, it is found that  for $B$ meson decays to
$J/\psi, \psi^\prime, \eta_c$ and $\eta_c^\prime$, the predicted
relative decay rates of these four states are approximately
compatible with experimental data. For instance, the theoretical
value \be
  \frac{{\mathrm{Br}}(B^0\rightarrow \eta_c K^0)}{{\mathrm{Br}} (B^0 \rightarrow
J/\psi K^0)} _{Th.}\approx
\bigl(\frac{f_{\eta_c}}{f_{J/\psi}}\bigr)^2
\cdot\bigl[\frac{F_1(m^2_{\eta_c})}{F_1(m^2_{J/\psi})}\bigr]^2
\cdot\bigl[\frac{m^2_B-m^2_{\eta_c}}{m^2_B-m^2_{J/\psi}}\bigr]^3
\approx 1.1 \times \bigl(\frac{f_{\eta_c}}{f_{J/\psi}}\bigr)^2
\approx 0.83-1.2,\label{r7}
 \ee
is compatible with the experimental value
   \be
 \frac{{\mathrm{Br}} (B^0\rightarrow \eta_c K^0)}
 {{\mathrm{Br}} (B^0 \rightarrow J/\psi K^0)}_{Ex.}\approx 1.0 .
 \label{r8}\ee
 Here $F_1$ is the $B \to K$ form factor, and the ratio of the squared
decay constants $(\frac{f_{\eta_c}}{f_{J/\psi}}\bigr)^2$ ranges
 from 0.75 to 1.1 in various models.

\subsection {$B \to\! \chi_{c0} K$, $B \to\! \chi_{c2} K$, $B \to\!
h_c K$ decays}

Studies of $B$ meson decays to P-wave charmonium states $B \to\!
\chi_{c0} K$, $B \to\! \chi_{c2} K$, and $B \to\! h_c K$
\cite{song2, song3} challenge the applicability of QCD
factorization for charmonium because of the appearance of
non-vanishing infrared divergences in these decays.

For these decays, there are no leading order contributions from
the $V-A$ currents in the effective 4-quark lagrangian, because
these P-wave charmonium states do not couple to the $V-A$
currents. According to \cite{BBNS,beneke} in QCD factorization all
nonfactorizable corrections are due to the vertex corrections and
spectator corrections, and other corrections are factorized into
the physical form factors and meson wave functions. Taking
nonfactorizable corrections into account, the decay amplitude for
$B\to\! \chi_{cJ}K(J=0,2)$ can be written as
 \begin{eqnarray}\label{amp2}
  i\mathcal{M} =\frac{G_F}{\sqrt{2}}\Bigl[V_{cb}
V_{cs}^* C_1-V_{tb} V_{ts}^* (C_4 + C_6) \Bigr]\times A,
\end{eqnarray}
where $C_i$ are the well known Wilson coefficients and $A$ is
given by \begin{eqnarray}\label{a} A=\frac{i 6
\mathcal{R}^{'}_1(0)} {\sqrt{\pi
M}}\cdot\frac{\alpha_s}{4\pi}\frac{C_F}{N_c} \Bigl(F_1\cdot f_I +
\frac{4\pi^2 f_K f_B}{N_c}  \cdot f_{II} \Bigr).
\end{eqnarray}
Here $R'_1(0)$ is the derivative of the radial wave function at
the origin of P-wave charmonia , $N_c$ is the number of colors,
$C_F=(N_c^2-1)/(2 N_c)$. The function $f_I$ is calculated from
vertex corrections, and $f_{II}$ is calculated from spectator
corrections. For ${}^3\!P_0$ charmonium state $\chi_{c0}$, the
infrared divergence from the vertex correction is found to be
 \begin{eqnarray}\label{fI1}
f_I\!=\!\frac{8 m_b z(1-z+\ln{z})}{(1-z)\sqrt{3
 z}}\ln{(\frac{\lambda^2}{m_b^2})}+finite\,\,terms,
  \end{eqnarray}
where $z\!=\!M^2/m_B^2 \approx 4m_c^2/m_b^2$, and $\lambda$ is the
gluon mass introduced to regularize the infrared divergences in
the vertex corrections. Similar infrared divergences can also be
found for the  $B \to\! \chi_{c2} K$ and $B \to\! h_c K$ decays.
The physical reason for the nonappearance of infrared divergences
for the S-wave charmonia $J/\psi$ and $\eta_c$ is based on the
fact that the virtual soft gluon in the vertex correction sees a
zero total color-charge of the charm quark plus the anti-charm
quark; whereas in the P-wave case the virtual soft gluon probes
not the color charge of the charm quark and the anti-charm quark
but the product of the color charge and the momentum of the charm
quark and anti-charm quark, and since the total contribution from
the charm quark plus anti-charm quark is nonzero (the relative
momentum between the charm quark and anti-charm quark is
essential) this leads to the appearance of infrared divergences
for the P-wave charmonia. Experimentally, $B \to\! \chi_{c0} K$
has a large rate, comparable to  $B \to\! \psi' K$ or $B \to\!
\chi_{c1} K$ which are factorizable with the leading order $V-A$
currents. This is certainly very puzzling in the QCD factorization
approach.

It is well known that the infrared divergences in the
\emph{inclusive} production and decay processes of a color-singlet
P-wave $c \bar c$ state can be removed by including contributions
 from the higher Fock states with color-octet $c \bar c$ pair (say
in S-wave) and soft gluon within the NRQCD factorization framework
\cite{bbl}. However,  the color-octet $c \bar c$ pair with
dynamical soft gluon can make contributions to the multi-body but
not two-body \emph{exclusive} decays.   As a result, the infrared
divergences encountered in \emph{exclusive} processes involving
charmonia may raise a new question to the QCD factorization and
NRQCD factorization in B exclusive decays. Further studies are
needed to seek the solution to remove the infrared divergences.

As for the leading order factorizable $B \to\! \chi_{c1} K$ decay,
the calculated rate is again too small \cite{song2}, as in other
leading order factorizable cases (e.g. $B \to\! J/\psi K$ and $B
\to\! \eta_c K$).

\subsection{  $B \to\! \psi(3770) K$ decay} The $\psi(3770)$ is the
$J^{PC}=1^{--}$ 1D-wave charmonium with a small admixture of 2S
component, and the mixing angle is estimated to be about
$\theta=-12^\circ$ \cite{mix}. Recently,  Belle \cite{belled} has
observed $\psi(3770)$ in the B meson decay
$B^+\rightarrow\psi(3770)K^+$ with a large branching ratio of
$(0.48\pm 0.11\pm 0.07)\times 10^{-3}$, which is comparable to
${\cal B}(B^+\rightarrow\psi'(3686)K^+)=(0.66\pm 0.06)\times
10^{-3}$ \cite{pdg}.

The $B \to\! \psi(3770) K$ decay rate is calculated with QCD
radiative corrections in QCD factorization \cite{gao} and it is
found that the 1D contribution is much smaller than 2S though both
1D and 2S components in the $\psi(3770)$ wave function can couple
to the leading order vector current and that a large mixing angle
$\theta=-26^\circ$ is required to fit the observed ratio
$\frac{Br(B\rightarrow\psi^{''}K)}{Br(B\rightarrow \psi^{'}K)}$.
The required large mixing angle is in contradiction with all other
estimates in charmonium phenomenology \cite{mix}. Moreover, the
calculated  $B \to\! \psi(3770) K$ decay rate is still smaller
than data by a factor of 8 even with the large mixing angle
included. In contrast, in the $B$ inclusive decays the D-wave
charmonium states are expected to have large rates due to the
color-octet mechanism regardless of the S-D mixing angle
\cite{yuanko}.

In summary, there are a number of problems in the description of
$B$ meson exclusive decays to charmonium states in the QCD
factorization approach. The decay rates for the S-wave states such
as $J/\psi$ and $\eta_c$ are too small to accommodate the
experimental data. For the exclusive decays to P-wave charmonia
$B\rightarrow \chi_{cJ} K$, it is found that (except $\chi_{c1}$)
there are infrared divergences arising from nonfactorizable vertex
corrections as well as logarithmic end-point singularities arising
 from nonfactorizable spectator interactions at leading-twist
order. The infrared divergences due to vertex corrections will
explicitly break down QCD factorization within the color-singlet
model for charmonia. Unlike in the inclusive decays where the
higher Fock states with color-octet $c \bar c$ pair and soft gluon
can make contributions to remove the infrared divergences, their
contributions can not be accommodated in the exclusive two body
decays.  As a result, the infrared divergences encountered in
exclusive processes involving charmonia may raise a new question
to the QCD factorization and NRQCD factorization in B exclusive
decays. For the D-wave state the observed $B \to\! \psi(3770) K$
decay rate would require a much larger S-D mixing angle than
conventional expected in charmonium phenomenology. All these
problems are likely to be related to the limitation of the
color-singlet $c\bar c$ pair description for charmonium in QCD
factorization. But how to solve these problems still remains an
open question. New theoretical considerations or ingredients in
QCD are needed.

\section{Perturbative QCD Approach based on $k_T$ Factorization}
\label{pqcd}\centerline{by C.D. L\"u, X.Q. Yu}\hspace{.8cm}

 Understanding non-leptonic $B$ meson decays is crucial
for testing the standard model, and also for uncovering the trace
of new physics. The simplest case is two-body non-leptonic $B$
meson decays, for which Bauer, Stech and Wirbel (BSW) proposed the
naive factorization assumption (FA) in their pioneering work
 \cite{BSW}. The technique to analyze hard exclusive hadronic
scattering was developed by Brodsky and Lepage  \cite{lb} based on
collinear factorization theorem in perturbative QCD (PQCD). A
modified framework based on $k_T$ factorization theorem was then
given in  \cite{BS,LS}, and extended to exclusive $B$ meson decays
in  \cite{LY1,CL,YL,CLY}. The infrared finiteness and gauge
invariance of $k_T$ factorization theorem was shown explicitly in
 \cite{NL}. Using this so-called PQCD approach, we have
investigated dynamics of non-leptonic $B$ meson decays
 \cite{KLS,LUY,KS}. Our observations are summarized as follows:
\begin{enumerate}
\item FA holds approximately for many charmless $B$ meson decays,
as our computation shows that non-factorizable contributions are
always negligible due to the cancellation between a pair of
non-factorizable diagrams.

\item Penguin amplitudes are enhanced, as the PQCD formalism
includes dynamics from the region, where the energy scale $\mu$
runs to $\sqrt{\bar\Lambda m_b}<m_b/2$, $\bar\Lambda\equiv
m_B-m_b$ being the $B$ meson and $b$ quark mass difference.

\item Annihilation diagrams contribute to large short-distance
strong phases through the $(S+P)(S-P)$ penguin operators.

\item The sign and magnitude of CP asymmetries in two-body
non-leptonic $B$ meson decays can be calculated, and we have
predicted relatively large CP asymmetries in the $B\to K^{(*)}\pi$
 \cite{KLS,Keum02} and $\pi\pi$ modes \cite{LUY,KS,Keum01}.
\end{enumerate}

\subsection{Formalism of PQCD Approach}

To develop the PQCD formalism for charmed $B$ meson decays, we
  investigated the $B\to D^{(*)}$ transition form factors in
the large recoil region of the $D^{(*)}$ meson  \cite{TLS2}.   The
$B\to D^{(*)}$ transition is more complicated than the $B\to\pi$
one, because it involves three scales: the $B$ meson mass $m_B$,
the $D^{(*)}$ meson mass $m_{D^{(*)}}$, and the heavy meson and
heavy quark mass difference, $\bar\Lambda=m_B-m_b\sim
m_{D^{(*)}}-m_c$ of order of the QCD scale $\Lambda_{\rm QCD}$,
$m_{D^{(*)}}$ ($m_c$) being the $D^{(*)}$ meson ($c$ quark) mass.
We have postulated the hierarchy of the three scales,
\begin{eqnarray}
m_B\gg m_{D^{(*)}}\gg\bar\Lambda\;, \label{hie}
\end{eqnarray}
which allows a consistent power expansion in $m_{D^{(*)}}/m_B$ and
in $\bar\Lambda/m_{D^{(*)}}$.

In the $B\to D^{(*)}$ transition , the initial state is
approximated by the $b\bar d$ component. The $b$ quark decays into
a $c$ quark and a virtual $W$ boson, which carries the momentum
$q$. Since the constituents are roughly on the mass shell, we have
the invariant masses $k_i^2\sim O(\bar\Lambda^2)$, $i=1$ and 2,
where $k_1$ ($k_2$) is the momentum of the spectator $\bar d$
quark in the $B$ ($D^{(*)}$) meson.
 The lowest-order diagrams contributing to the
$B\to D^{(*)}$ form factors contain a hard gluon exchange between
the $b$ or $c$ quark and the $\bar d$ quark. The $\bar d$ quark
undergoes scattering in order to catch up with the $c$ quark,
forming a $D^{(*)}$ meson. The exchanged gluon is off-shell by
\begin{eqnarray}
(k_1-k_2)^2\sim -\frac{m_B}{m_{D^{(*)}}}\bar\Lambda^2\;,
\label{hs1}
\end{eqnarray}
which has been identified as the characteristic scale of the hard
kernels. Under Eq.~(\ref{hie}), we have $m_B/m_D^{(*)}\gg 1$, and
the hard kernels are calculable in perturbation theory. It has
been found that the applicability of PQCD to the $B\to D^{(*)}$
transition at large recoil is marginal for the physical masses
$m_B$ and $m_{D^{(*)}}$  \cite{TLS2}.

 The form factors are then expressed as the convolution of the
hard kernels $H$ with the $B$ and $D^{(*)}$ meson wave functions
in $k_T$ factorization theorem,
\begin{eqnarray}
F^{BD^{(*)}}(q^2)=\int dx_1dx_2 d^2b_1 d^2b_2
\phi_B(x_1,b_1)H(x_1,x_2,b_1,b_2)\phi_{D^{(*)}}(x_2,b_2)\;.
\label{bds}
\end{eqnarray}
The $D^{(*)}$ meson wave function contains a Sudakov factor
arising from $k_T$ resummation, which sums the large double
logarithms $\alpha_s\ln^2(m_Bb_2)$ to all orders. The $B$ meson
wave function also contains such a Sudakov factor, whose effect is
negligible because a $B$ meson is dominated by soft dynamics. The
hard kernels involve a Sudakov factor from threshold resummation,
which sums the large double logarithm $\alpha_s\ln^2 x_1$ or
$\alpha_s\ln^2x_2$ to all orders. This factor modifies the
end-point behavior of the $B$ and $D^{(*)}$ meson wave functions
effectively, rendering them diminish faster in the small $x_{1,2}$
region.


It has been pointed out that if evaluating the hard part with a
hard gluon in collinear factorization theorem, end-point
singularities appear  \cite{SHB}. These singularities imply the
breakdown of collinear factorization, and $k_T$ factorization
becomes more appropriate. Once the parton transverse momenta $k_T$
are taken into account, a dynamical effect, the so-called Sudakov
suppression, favors the configuration in which $k_T$ is not small
 \cite{LS}. The end-point singularities then disappear. The Sudakov
factor suppresses the large $b$ region, where the quark and
anti-quark are separated by a large transverse distance and the
color shielding is not so effective. It also suppresses the $x\sim
1$ region, where a quark carries all of the meson momentum, and
intends to emit real gluons in hard scattering.

\subsection{$B\to D^{(*)}\pi$ IN PQCD}

In this subsection we take the $B\to D\pi$ decays as an example of
the PQCD analysis. In the PQCD framework based on $k_T$
factorization theorem, an amplitude is expressed as the
convolution of hard $b$ quark decay kernels with meson wave
functions in both the longitudinal momentum fractions and the
transverse momenta of partons. Our PQCD formulas are derived up to
leading-order in $\alpha_s$, to leading power in $m_D/m_B$ and in
$\bar\Lambda/m_D$, and to leading double-logarithm resummations.

\begin{table}
\begin{center}
\begin{tabular}{|l| c| c|c|c| }
\hline Quantities & $C_D=0.6$ & $C_D=0.8$ & $C_D=1.0$ & Data \\
\hline
$B(\bar B^0\to D^+\pi^-)$ & 2.37 & 2.74 & 3.13 & $3.0\pm 0.4$ \\
$B(\bar B^0\to D^0\pi^0)$ & 0.26 & 0.25 & 0.24 & $0.29 \pm 0.05$\\
$B(B^-\to D^0\pi^-)$ & 4.96 & 5.43 & 5.91 & $5.3\pm 0.5$ \\
\hline \hline
Decay Modes & $C_{D^*}=0.5$ & $C_{D^*}=0.7$ & $C_{D^*}=0.9$ & Data \\
\hline
$B(\bar B^0\to D^{*+}\pi^-)$ & 2.16 & 2.51 & 2.88 & $2.76\pm 0.21$ \\
$B(\bar B^0\to D^{*0}\pi^0)$ & 0.29 & 0.28 & 0.27 & $0.25 \pm 0.07$\\
$B(B^-\to D^{*0}\pi^-)$ & 4.79 & 5.26 & 5.75 & $4.60\pm 0.40$ \\
\hline
\end{tabular}
\caption{Predicted $B\to D^{(*)}\pi$ decay  branching ratios in
units of $10^{-3}$.} \label{dp}
\end{center}
\end{table}

The predicted branching ratios in Table~\ref{dp} are in agreement
with the averaged experimental data  \cite{BelleC,CLEOC,Bab}. We
extract the effective parameters $a_1$ and $a_2$ from our PQCD
calculations. That is, our $a_1$ and $a_2$ do not only contain the
non-factorizable amplitudes as in generalized FA, but the small
annihilation amplitudes, which was first discussed in
\cite{GKKP}. We obtain the ratio $|a_2/a_1|\sim 0.43$ with $10\%$
uncertainty and the phase of $a_2$ relative to $a_1$ about
$Arg(a_2/a_1)\sim -42^\circ$.   Note that the experimental data do
not fix the sign of the relative phases. The PQCD calculation
indicates that $Arg(a_2/a_1)$ should be located in the fourth
quadrant. It is evident that the short-distance strong phase from
the color-suppressed non-factorizable amplitude is already
sufficient to account for the isospin triangle formed by the $B\to
D\pi$ modes.  From the viewpoint of PQCD, this strong phase is of
short distance, and produced from the non-pinched singularity of
the hard kernel. Certainly, under the current experimental and
theoretical uncertainties, there is still room for long-distance
phases from final-state interaction.

The PQCD predictions for the $B\to D^*\pi$ decay branching ratios
in Table~\ref{dp} are also consistent with the data \cite{pdg}.
Since $m_{D^*}$ and $\phi_{D^*}$ are slightly different from
$m_{D}$ and $\phi_{D}$, respectively, the results are close to
those  for $B\to D\pi$. Similarly, the ratio $|a_2/a_1|$ and the
relative phase $Arg(a_2/a_1)$ are also close to those associated
with the $B\to D\pi$ decays. We obtain the ratio $|a_2/a_1|\sim
0.47$ with $10\%$ uncertainty and the relative phase about
$Arg(a_2/a_1)\sim -41^\circ$. The $B\to D^{(*)} \eta^{(\prime)} $
decays are similar to the $B\to D^{(*)} \pi^0 $ decays except the
$\eta$ -$\eta'$ mixing part \cite{deta}. The predicted branching
ratios for these decays are shown in table \ref{pqcdr}, with
uncertainties only from the $\eta$ -$\eta'$ mixing parameter
$\theta$.

\begin{table}
\begin{center} \caption{PQCD predictions with one angle mixing formalism (I) and
two angle mixing formalism (II) and experimental data (in units of
$10^{-4}$)
               of the $B^0\to \bar D^{(*)0} \eta^{(\prime)} $ branching ratios.}
\begin{tabular}{|l| l| c| c|c|c|c|}
\hline {Decay mode}             &     PQCD (I)  & PQCD (II) &
Belle & BaBar &PDG
 \\ \hline
 $B^0\to \bar D^{0}\eta^{\prime}$ &  $1.7\sim 2.3$  &  $2.2\sim 2.6$  &  - &- &  $ <9.4 $\\
 ${ B}^0\to \bar D^{0}\eta $ &  $2.4\sim 3.0$ &  $2.6\sim 3.2$ & $1.4^{+0.6}_{-0.5}$
 &$2.41 \pm 0.50 $ & \\
    ${ B}^0\to \bar D^{*0}\eta'$ & $2.0 \sim 2.7 $ &  $2.6\sim 3.2$&  & - &  $ < 14 $\\
 ${ B}^0\to \bar D^{*0}\eta $ &  $2.8\sim 3.5$ &  $3.1\sim 3.8$& $2.0^{+1.0}_{-0.9}$ &- &\\
     \hline
\end{tabular}
\label{pqcdr}
\end{center}
\end{table}

\subsection{$B_s \to \pi^+\pi^-$ decay IN PQCD}

We also calculate the decay rate and CP asymmetry of the $B_s \to
\pi^+\pi^-$ decay in perturbative QCD approach with Sudakov
resummation. Since none of the quarks in final states is the same
as those of the initial $B_s$ meson, this decay can occur only via
annihilation diagrams in the standard model. Besides the
current-current operators, the contributions from the QCD and
electroweak penguin operators are also taken into account. We find
that (a) the branching ratio is about $4 \times 10^{-7}$; (b) the
penguin diagrams dominate the total contribution; and (c) the
direct CP asymmetry is small in size: no more than $3\% $; but the
mixing-induced CP asymmetry can be as large as ten percent
testable in the near future LHC-b experiments and BTeV experiments
at Fermilab. This small branching ratio, predicted in the SM, make
it sensitive to possible new physics contribution.

\section{B meson decays and the CKM angles $\alpha, \beta$ and $\gamma$}
\label{ckm}   \centerline{by Z.J. Xiao}\hspace{0.8cm}

For discussion on isospin and SU(3) relations in charmless B
decays and possible new type of electroweak penguin or SU(3)
breaking effects of strong phases, as well as large direct CP
violation in charmless B decays, it is suggested to see ref.
\cite{WZ1}. As a summary, we observe that in the case of SU(3)
limits and also the case with SU(3) breaking only in amplitudes,
the fitting results lead to an unexpected large ratio between two
isospin amplitudes $a^{c}_{3/2}/a^{u}_{3/2}$, which is about an
order of magnitude larger than the SM prediction.  The results are
found to be insensitive to the weak phase $\gamma$.  By including
SU(3) breaking effects on the strong phases, one is able to obtain
a consistent fit to the current data within the SM, which implies
that the SU(3) breaking effect on strong phases may play an
important role in understanding the observed charmless hadronic B
decay modes $B\to \pi \pi$ and $\pi K$. It is possible to test
those breaking effects in the near future from more precise
measurements of direct CP violation in B factories.

The direct CP violation with SU(3) symmetry breaking of strong
phases can be as large as follows:
\begin{eqnarray*}
 & & A_{CP}(\pi^{+}\pi^{-}) \simeq  0.5 , \qquad  A_{CP}(\pi^{0}\pi^{0}) \simeq  0.2 ,
 \qquad  A_{CP}(\pi^{+} K^-) \simeq -0.1, \\
 & &  A_{CP}(\pi^{0}\bar K^{0}) \simeq  -0.1 , \qquad A_{CP}(\pi^{0} K^-) \simeq -0.0,
 \qquad A_{CP}(\pi^{-} \bar K^0) \simeq  0.1.
\end{eqnarray*}

For CP violation in $B\rightarrow \phi K_S$ and possible new
physics models, see ref. \cite{WZ2} and references therein.

 One of the most important tasks in B factory
experiments is to measure the CKM angles $\alpha, \beta$ and
$\gamma$ \cite{0304132}, the three inner angles of the unitarity
triangle defined in the complex $(\bar{\rho},\bar{\eta})$ plane.
At present, $B \to \pi\pi, K\pi$ and many other interesting decays
have been measured with good precision in B factory experiments.
These modes together with the CP asymmetries of $B \to J/\Psi K_S,
\rho\rho,  \rho^\pm \pi^\mp, etc$ will allow us to determine the
CKM angles $\alpha$, $\beta$ and $\gamma$. The isospin symmetry of
strong interaction and $SU(3)$ flavor symmetry are frequently used
to deal with "unknown" hadronic matrix elements, and in some cases
also plausible dynamical assumptions have to be made.

 From a global fit done by CKM fitter group  \cite{0406184} the
currently allowed ranges of these three angles at $95\%$ C.L. are
\begin{eqnarray}
78^\circ \leq \alpha \leq 122^\circ, \quad 21^\circ \leq \beta
\leq 27^\circ , \quad 38^\circ \leq \gamma \leq 80^\circ
\end{eqnarray}

The angle $\beta$ has already been well measured using the
time-dependent CP asymmetries in the $B_d^0 \to J/\Psi K_S$ and
related decays. The world average as given by Heavy Flavor
Averaging Group  \cite{hfag04} is
\begin{eqnarray}
\sin{2\beta} &=& 0.739 \pm 0.048, \ \  all \ \ charmonium, \\
\nonumber
     &=& 0.692 \pm 0.045, \ \  all \ \ modes
\end{eqnarray}
which leads to the bounds on the angle $\beta$:
\begin{eqnarray}
\beta &=& \left ( 23.8^{+2.2}_{-2.0} \right )^\circ  \bigvee \left
(  66.2 ^{-2.2}_{+2.0} \right )^\circ,\ \  all \ \ charmonium,
\\ \nonumber
&=& \left ( 21.9^{+1.9}_{-1.7} \right )^\circ  \bigvee \left (
68.1 ^{-1.9}_{+1.7} \right )^\circ,\ \  all \ \ modes
 \label{eq:b-beta}
\end{eqnarray}
The twofold ambiguity has been resolved by BaBar's new measurement
of $\cos{2\beta}$: a negative $\cos{2\beta}$ is excluded at
$89\%$C.L.  \cite{0406082ex}. We therefore believe that
$\beta=24^\circ \pm 2^\circ$ is a reliable measurement and can be
used as an experimental input in the effort to determine other two
angles $\alpha$ and $\gamma$.

In the SM, the penguin-dominated $B \to \phi K_s$ decay should
measure the same $\sin{2\beta}$ as $B \to J/\Psi K_S $ decay.
However, a $3.5\sigma$ deviation from the SM prediction is
observed by Belle collaboration. The most recent measurements give
 \cite{phiks}
\begin{eqnarray}
\sin{2\beta}|_{\phi K_s} = \left\{ \begin{array}{ll }
-0.96 \pm 0.50 ^{+0.09}_{-0.11} & (Belle) \\
+ 0.47 \pm 0.34 ^{+0.08}_{-0.06} & (BaBar)
\end{array}  \right.
\end{eqnarray}
Although it is too early to draw any conclusion from above primary
measurements, many works have been done to interpret such
deviation as a hint of new physics contributions to the quark
level $b \to ss\bar{s}$ decays.

The CKM angle $\alpha$ can be extracted through the CP violation
in the tree-dominated $B \to \pi\pi, \rho \pi$ and $B \to \rho
\rho$ decays, The $B \to \pi^+ \pi^-$ decay currently plays most
important rule in constraining $\alpha$. The isospin symmetry can
be used to eliminate the penguin pollution, but a
model-independent isospin analysis of $B \to \pi\pi$ decays
requires the knowledge of the three amplitudes $A^{+-}, A^{00},
A^{+0}$ and their charge conjugates $\bar{A}^{ij}$, At present,
the only missing pieces are $A^{00}$ and $\bar{A}^{00}$. It is
possible that a full isospin analysis will be done, and $\alpha$
extracted cleanly, by the summer of 2005.

Theoretically, the CP asymmetry parameter $C_{\pi\pi}$ and
$S_{\pi\pi}$ depend on $\alpha$, the strong phase $\delta$ and the
ratio $r=|P/T|$:
\begin{eqnarray}
S_{\pi\pi} &=& \frac{\sin{2\alpha} - 2 r \cos{\delta }
\sin{\alpha}}{ 1+ r^2 - 2r\cos{\delta}\cos{\alpha}}, \quad
C_{\pi\pi}  \frac{2 r \sin{\delta} \sin{\alpha}}{ 1+ r^2 -
2r\cos{\delta}\cos{\alpha}}.
\end{eqnarray}

In ref. \cite{alpha}, taking  $S_{\pi \pi}=-0.49 \pm 0.27$ and
$C_{\pi\pi}=0.51\pm 0.19$ as experimental input, we found the
allowed ranges: (a) $76^\circ \leq \alpha \leq 135^\circ$ for
$r=0.3 \pm 0.1 $; and (b) $117^\circ \leq \alpha \leq 135^\circ$
and $-160^\circ \leq  \delta \leq -132^\circ$ if we also take the
measured branching ratios of $B \to \pi^+ \pi^-, \pi^+ \pi^0$ and
$K^0 \pi^+$ decays into account.

In ref. \cite{g04}, by setting $f_K/f_\pi$ as the $SU(3)$ flavor
symmetry breaking factor,  and using the measured branching ratios
of $B \to \pi^+\pi^-$ and $K\pi$ decays and the new average
$S_{\pi\pi} = -0.74 \pm 0.16$ and $C_{\pi\pi} = -0.46 \pm 0.13$
 \cite{hfag04} as input, Gronau found the allowed range, $\alpha =
(104\pm 18)^\circ$  \cite{g04}. From currently available
measurements of $B \to \rho \rho$ decays, the range $19^\circ \leq
\alpha \leq 71^\circ$ is excluded at $90\%$C.L.  \cite{g04}. From
the measured time-dependent CP asymmetries in $B \to \rho^\pm
\pi^\mp$, one found that the allowed range is $\alpha = 93^\circ
\pm 17^\circ$ ($\alpha = 102^\circ \pm 11^\circ$) if only the
BaBar (Belle) measurements are taking into account.

By using the SM relation $\alpha + \beta + \gamma=180^\circ$, the
above bounds on $\alpha$ can be translated into bounds on $\gamma$
for fixed value of $\beta =24^\circ \pm 2 ^\circ$. Of course, many
strategies to determine angle $\gamma$ directly from B meson
decays have been proposed recently
\cite{0304132,gamma,xiao02,pi04}, for example,
\begin{itemize}
\item Isospin symmetry plus the $B \to K \pi$ observables allow
one to constrain the angle $\gamma$. In ref. \cite{xiao02}, for
example, the region $\gamma \sim 90^\circ$ was excluded by
considering a "mixed" system of $B^+ \to \pi^+ K^0$ and $B_d^0 \to
\pi^0 K^0$ decays.

\item U-spin symmetry plus the measurements of $B_d \to \pi\pi$
and $B_s \to KK$, , $B_d \to J/\Psi K_S$ and $B_s \to D_s^+
D_s^-$, or $B_s \to D_s^{(*)\pm } K^\mp $ and $B_d \to D^{(*)\pm}
\pi^\mp$, etc.

\item to determine $\alpha$ through the measurements of the
branching ratios of six $B \to D K$ decays.

\end{itemize}

\section{ Instantaneous Bethe-Salpeter Equation and Its Exact Solutions}
\label{bs}      \centerline{by C.H. Chang}\hspace{0.8cm}

We propose an approach to solve a Bethe-Salpeter (BS) equation
exactly without approximation if its kernel exactly is
instantaneous, and take positronium as an example to illustrate
the general features of the approach and the obtained solutions
under it \cite{BS-in}.

The key step for the approach is to start with the BS equation to
derive out a set of coupled and well-determined integration
equations for the components of the BS wave function without any
approximation as a linear eigenvalue problem. The coupled
equations may be solved analytically and/or numerically depending
on the specific kernel. If there is no analytic solution, in
principle, the coupled equations can be also solved numerically
under a controlled (requested) accuracy.

To solve the coupled and well-determined equations for
positronium, a numerical method, which is to expand the equations
(the eigenfunctions and the kernel) for the BS wave function
components in terms of the relevant Schr\"odinger equation
solutions with suitable truncation and then alternatively to solve
the truncated matrix equation by making it diagonal, is applied,
and accurate enough solutions for requests (the accuracy depends
on the truncation mainly) are obtained. The exact solutions
present precise and substantial corrections to those of the
corresponding Schr\"odinger equation, which start from the order
${\cal O}(v^1)$ ($v$ is the relative velocity) for eigenfunctions,
the order ${\cal O}(v^2)$ for eigenvalues and the mixing between
$S$ ($P$) and $D$ ($F$) components in $J^{PC}=1^{--}$
($J^{PC}=2^{++}$) states etc. From the lessons, we point out that
such corrections are important in the effective theories such as
NRQCD and NRQED when we consider the relativistic corrections
${\cal O}(v)\sim {\cal O}(\frac{1}{M})$. Namely {\bf one cannot
`forget' the same order corrections involved in the solutions of
the bound states accordingly.}

Moreover, we also point out that there is a questionable step in
the classical derivations for an instantaneous BS equation, if one
is pursuing the exact solutions without approximation. Namely it
has confused the differences between the instantaneous BS equation
and Breit equation (more precise illustration in \cite{brei-e}).

\section{New Physics Effects in B Decays}
\label{new}    \centerline{by X.G. He and C.S. Huang and Z.J.
Xiao}\hspace{0.8cm}

\subsection{Time Dependent CP Violation in $B\to K^* \gamma$ in
SUSY}

  In this work we study  time
dependent CP asymmetries in $B\to K^* \gamma \to \pi K_S\gamma$
and $B \to \phi K_S$ using constraints from $B\to X_S \gamma$ in
SUSY model with large Left-Right (or Right-Left) squark mixing
induced gluonic dipole interaction. Since this work was reported
at the meeting, new results on $B$ decays have been reported
\cite{expcpasy, hfag,cpgamma,babar}, in particular the direct CP
asymmetry $A_{CP}$ in $\bar B^0 \to K^- \pi^+$ of \cite{expcpasy}
$-0.114\pm 0.020$ measured by Babar and Belle, we include the
constraint from $A_{CP}(K^-\pi^+)$ in our study \cite{hly}.

In the SM, the Hamiltonian for the $B$ decays to be considered is
well known which is of the form \cite{sm},
\begin{eqnarray}
H=\frac{G_F }{ \sqrt{2}} [ V_{ub} V^*_{us}(c_1 O_1 +c_2 O_2)-
\sum^{12}_{i=3}V_{jb}V_{js}^*c^j_iO_i],
\end{eqnarray}
where $V_{ij}$ are the CKM matrix elements. $c_i$ are the Wilson
coefficients for the operators $O_i$ which have been evaluated in
different schemes. Values from NDR scheme will be used \cite{sm}.
We will not display the full sets of $O_i$ and $c_i$ here, but
only give the definitions of the gluonic and photonic dipole
operators $O_{11}$ and $O_{12}$ for the convenience of later
discussions. They are given by
\begin{eqnarray}
O_{11} &=& \frac{g_s}{ 8\pi^2} \bar s \sigma_{\mu\nu}
G^{\mu\nu}_a T^a[m_b(1+\gamma_5)+m_s(1-\gamma_5)]b,\nonumber\\
O_{12} &=& \frac{e}{ 8\pi^2} \bar s \sigma_{\mu\nu} F^{\mu\nu}
[m_b(1+\gamma_5)+m_s(1-\gamma_5)]b,
\end{eqnarray}
where $T^a$ is the color SU(3) generator normalized to $Tr(T^a
T^b) = \delta^{ab}/2$. $G_{\mu\nu}$ and $F_{\mu\nu}$ are the gluon
and photon field strengths. In the SM \cite{sm,sm1,kagan}
$c_{11}=-0.151$ and $c_{12} = -0.318$.

When going beyond the SM, there are modifications. In SUSY models,
exchanges of gluino and squark with Left-Right squark mixing can
generate a large contribution to $c_{11,12}$ at one loop level
\cite{susydipole,massinsertion} since their interactions are
strong couplings in strength and also enhanced by a factor of the
ratio of gluino mass to the $b$ quark mass \cite{pheno}. We will
concentrate on the effects of this interaction. In general
exchange of squarks and gluinos can generate non-zero $c_{11,12}$
for dipole operators with $1+\gamma_5$, as well as with non-zero
$c'_{11,12}$ for dipole operators with $1-\gamma_5$.

The Wilson coefficient $c^{susy}_{11,12}$ for SUSY contribution
obtained in the mass insertion approximation is given by, for the
case with $1+\gamma_5$ \cite{massinsertion},

\begin{eqnarray}
&&c_{11}^{susy}(m_{\tilde g}) = \frac{\sqrt{2}\pi
\alpha_s(m_{\tilde g})}{ G_F m^2_{\tilde g}}
\frac{\delta_{LR}^{bs}}{ V_{tb}V_{ts}^*}\frac{m_{\tilde g}}{ m_b}
G_0(x_{gq}),\nonumber\\
&&c_{12}^{susy}(m_{\tilde g}) = \frac{\sqrt{2}\pi
\alpha_s(m_{\tilde g})}{ G_Fm^2_{\tilde g}}
\frac{\delta_{LR}^{bs}}{ V_{tb}V_{ts}^*}\frac{m_{\tilde g}}{ m_b}
F_0(x_{gq}),\nonumber\\
&&G_0(x) = \frac{x[22-20x-2x^2+(16x-x^2+9)\ln(x)] }{ 3(1-x)^4},
\nonumber\\
&&F_0(x) = -\frac{4x[1+4x-5x^2+(4x +2x^2)\ln(x)] }{ 9(1-x)^4},
\label{dipole}
\end{eqnarray}
where $\delta_{LR}^{bs}$ parameterizes  the mixing of left and
right squarks, $x_{gq} = m^2_{\tilde g}/m^2_{\tilde q}$ is the
ratio of gluino mass $m_{\tilde g}$ and squark mass $m_{\tilde
q}$. The Wilson coefficients $c^{'susy}_{11,12}$ for the case with
$1-\gamma_5$ can be easily obtained by replacing the Left-Right
mixing parameter $\delta_{LR}^{bs}$ by the Right-Left mixing
parameter $\delta^{bs}_{RL}$.

 From the expressions in eq.(\ref{dipole}), one can see that the
SUSY contributions are proportional to $m_{\tilde g}$. If
$m_{\tilde g}$ is of order a few hundred GeV, there is an
enhancement factor of $(m_{\tilde g}/m_b) (m_W^2/m_{\tilde g}^2)$
for the SUSY dipole interactions. In this case even a small
$\delta_{LR, RL}^{bs}$, which can easily satisfy constraints from
$B-\bar B$ mixing and other data, can have a large effect on rare
$B$ decays.

We first consider constraint on the SUSY parameters
$\delta^{bs}_{LR,RL}$ from $B\to X_S \gamma$. The branching ratio
of this process has been measured to a good precision with
\cite{hfag} $(3.54^{+0.30}_{-0.28})\times 10^{-4}$. Although
experimentally CP asymmetry in $B\to X_S\gamma$ has not been
established, there are constraints from experiments with
\cite{cpgamma} $0.005 \pm 0.036$. We follow ref.\cite{kagan} to
carry out the evaluation of the branching ratio and CP asymmetry
and use them to constrain the parameters.

\begin{figure}[htb]
\begin{center}
\includegraphics[width=6cm]{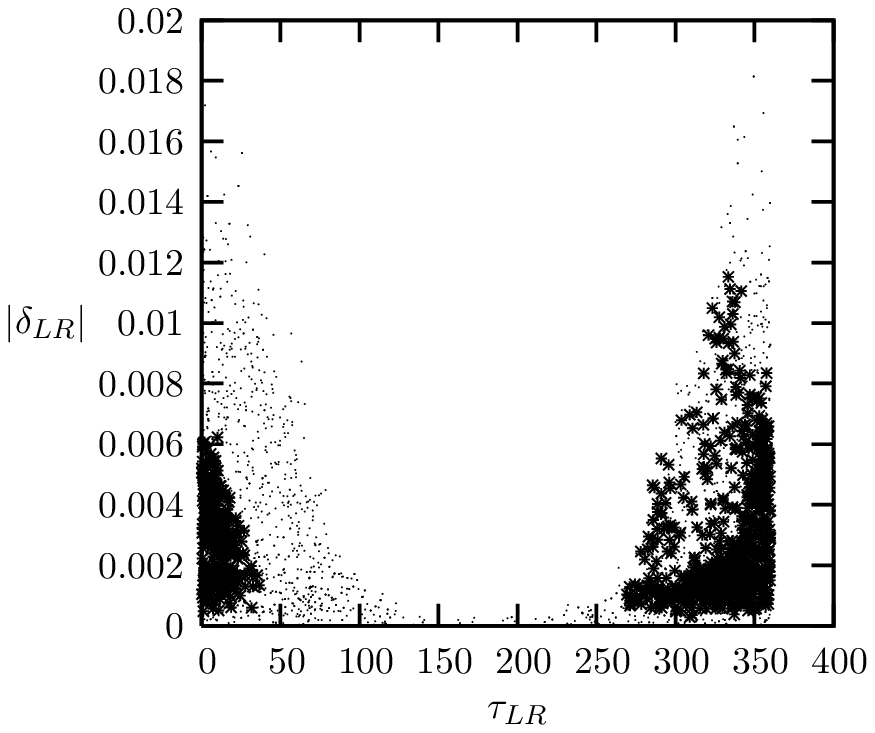}
\includegraphics[width=6cm]{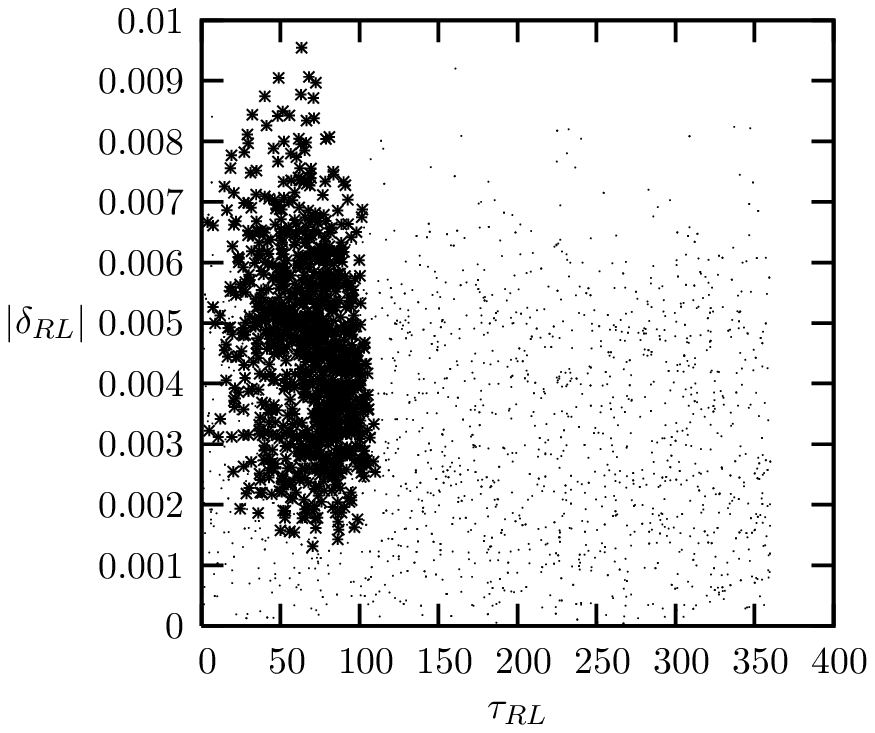}
\end{center}
\caption{The one $\sigma$ allowed ranges for the SUSY parameters
$|\delta_{LR,RL}^{bs}|$ and the phase $\tau$ taking $m_{\tilde
g}=300$GeV and $m_{\tilde q}$ in the range $100 \sim 1000$ GeV.
The light-dark doted areas are the allowed parameter spaces from
$BR(B\rightarrow \, X_s \gamma)$ and $A_{CP}(B\to X_S\gamma)$
constraints. The dark doted areas are allowed ranges by $A_{CP}(
\bar B^0 \to K^- \pi^+)$ constraint. Figure 1a (on the left)
 and Figure 1b (on the right) are for the
dipole operators with $1+\gamma_5$ and $1-\gamma_5$,
respectively.} \label{f2}
\end{figure}

In Figure \ref{f2} we show the allowed ranges for the absolute
value of $\delta^{bs}_{LR}$ and its phase $\tau$ for $m_{\tilde
g}=300$GeV and $m_{\tilde q}$ in the range $100 \sim 1000$GeV at
the one $\sigma$ level.  We find that the constraints from
$BR(B\to X_S\gamma)$ and $A_{CP}(B\to X_S\gamma)$ are similar.
Using the allowed parameters, one can obtain the allowed $c_{11}$
through eq.\ref{dipole} and to study implications for other rare
$B$ decays.


The recently measured CP asymmetry $A_{CP}(\bar B^0 \to K^-
\pi^+)$ can be reproduced and the SUSY parameters can be further
constrained. We follow ref.\cite{bbns} to calculate the SUSY
dipole contribution to $B\to K \pi$. In our numerical analysis we
take the CKM parameters to be known, with the standard
parametrization $s_{12}=0.2243$, $s_{23}=0.0413$, $s_{13}=0.0037$,
$\delta_{13}=1.05$, which is the central value given by the
Particle Data Group \cite{pdg}. With the SM amplitudes obtained
and the default values for the hadronic parameters used in Ref.
\cite{bbns}, we obtain the CP asymmetry $A_{CP}(\bar B^0 \to K^-
\pi^+)$ in the SM to be 0.15. This is different in sign with the
experimental value. When SUSY dipole interactions are included the
experimental value can be reproduced. For example with
$m_{\tilde{g}}=m_{\tilde{q}}=300$GeV, $\delta_{LR}=2.62\times
10^{-3}e^{0.238i}$, $\delta_{RL}=4.31\times 10^{-3}e^{1.007i}$ the
asymmetry $A_{CP}(\bar B^0 \to K^- \pi^+)$ is approximately
-0.114. With the same set of SUSY parameters, we have $Br(B\to
X_s\gamma)=3.48\times 10^{-4}$, $A_{CP}(B\to X_S \gamma)=0.016$.
It is clear that the CP asymmetry $A_{CP}(\bar B^0\to K^- \pi^+)$
can be brought to be in agreement with data at one sigma level
when SUSY gluonic dipole interactions are included.

To see how the CP asymmetry provides stringent constraint on the
SUSY flavor changing parameters, we show in Figure \ref{f2} the
parameter space allowed from $A_{CP}(\bar B^0\to K^- \pi^+)$ (the
dark doted areas) on top of the allowed ranges by $B\to X_S
\gamma$ constraint alone at the one $\sigma$ level. We see that
the CP asymmetry in $\bar B^0 \to K^- \pi^+$ considerably reduces
the allowed parameter space.

We now study time dependent CP asymmetries in $B\to K^*\gamma\to
\pi^0 K_S \gamma$ and $B\to \phi K_S$. There are two CP violating
parameters $A_f$ and $S_f$ which can be measured in time dependent
decays of $B$ and $\bar B$ produced at $e^+e^-$ colliders at the
$\Upsilon(4S)$ resonance, $A^{CP}(t) = A_f \cos(\Delta t \Delta
m_B) + S_f \sin(\Delta t \Delta m_B)$.

For $\bar B^0 \to \bar K^{*0} \gamma \to \pi^0 K_S \gamma$ and
$B^0 \to K^{*0}\gamma \to \pi^0 K_S \gamma$, we obtain
\cite{kagan,pheno,atwood}

\begin{eqnarray}
S_{K^*\gamma} &=&
  -2\frac{Im[(q_B/p_B)(c_{12}c'_{12})]}{|c_{12}|^2+|c'_{12}|^2}.
\end{eqnarray}
To the leading order $A_{K^*\gamma}$ is the same as $A_{CP}(B\to
X_S \gamma)$. Note that the hadronic matrix element $<K^*|\bar s
\sigma^{\mu\nu} (1\pm \gamma_5) b|B>$ does not appear making the
calculation simple and reliable. In order to have a non-zero
$S_{K^*\gamma}$ both $c_{12}$ and $c'_{12}$ cannot be zero.

In the SM the asymmetries $A_{K^*\gamma}$ and $S_{K^*\gamma}$ are
predicted to be small with \cite{kagan, atwood}
$A^{SM}_{K^*\gamma} \approx 0.5\%$, $S^{SM}_{K^*\gamma} \approx
3\%$. With SUSY gluonic dipole interaction, the predictions for
these CP asymmetries can be changed dramatically \cite{pheno}.
With the constraints obtained previously, we find that the
parameter $q_B/p_B$ is not affected very much compared with the SM
calculation. To a good approximation $q_B/p_B =e^{-2i\beta}$.

A Large gluonic dipole interaction also has a big impact on $B\to
\phi K_S$ decays \cite{phiK}. In the SM, $A_{\phi K_s}$ is
predicted to be very small and $S_{\phi K_S}$ is predicted to be
the same as $S_{J/\psi K_S}=\sin (2\beta)$. With SUSY gluonic
dipole contribution, the decay amplitude for $B\to \phi K_S$ will
be changed and the predicted value for both $A_{\phi K_S}$ and
$S_{\phi K_S}$ can be very different from those in the SM
\cite{phiK}. We again use QCD factorization \cite{bbns} to
evaluate the amplitude.


\begin{figure}[htb]
\begin{center}
\includegraphics[width=6cm]{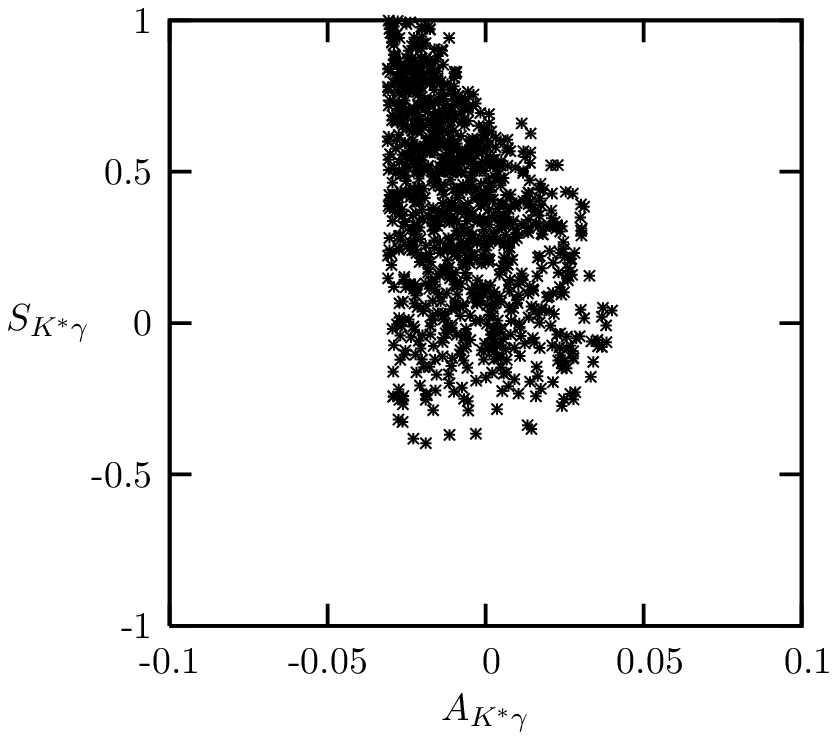}
\includegraphics[width=6cm]{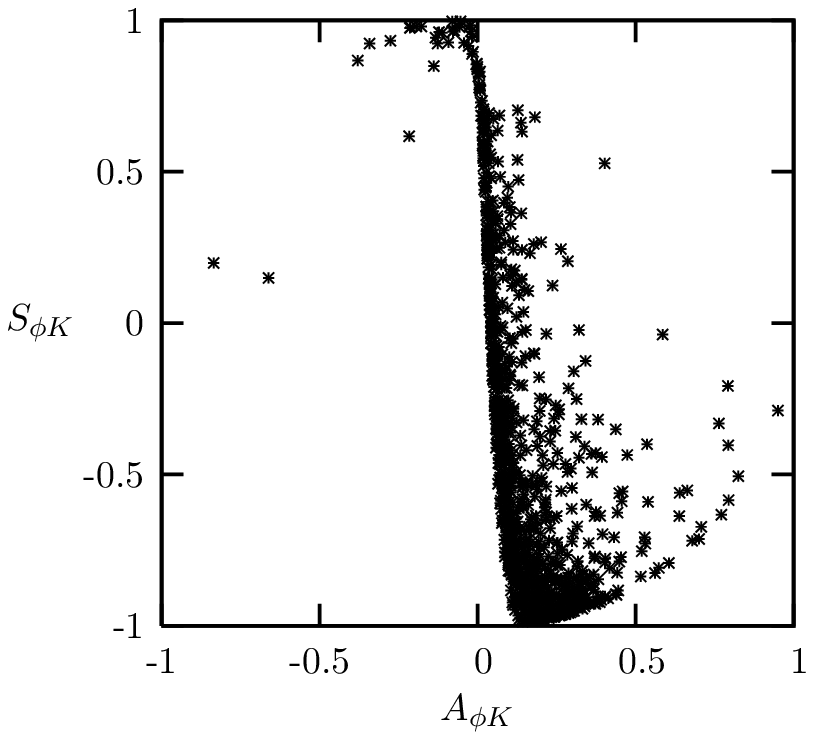}
\end{center}
\caption{The allowed time dependent CP asymmetries in $ \bar B^0
\to K^*\gamma\to K_S \pi^0 \gamma$ and $\bar B^0\to \phi K_S$. }
\label{f3}
\end{figure}

The results are shown in Figure 3. The current values of
$S_{K^*\gamma}$ and $A_{K^*\gamma}$ from Babar (Belle) are
\cite{kgamma}: $0.57\pm 0.32\pm 009(-0.00\pm 0.38)$, $0.25\pm
0.63\pm0.14(-0.79^{+0.63}_{-0.50}\pm 0.09)$, respectively. From
Figure 3, we see that the allowed ranges can cover the central
values of $S_{K^*\gamma}$ from Babar and Bell, but it is not
possible to obtain the central value of $A_{K^*\gamma}$ by Belle.
Future improved data can further restrict the parameter space.
Both Babar and Belle have also measured \cite{ss} $A_{CP}(B^- \to
K^{*-}\gamma)$ with ranges $-0.074 \sim 0.049$(Babar) and
$-0.015\pm 0.044\pm 0.012$(Belle). In the model we are
considering, the CP asymmetries $A_{K^*\gamma}$ and $A_{CP}(B^-
\to K^{*-}\gamma)$ are the same. The results for the charged $B$
CP asymmetry is consistent with data.

The time dependent asymmetry in $B\to \phi K_S$ is a very good
test of CP violation in the SM. Experimental measurements have not
converged with the current values of Babar (Belle) given by
\cite{kphi,babar} $0.00\pm0.23\pm0.05(0.08\pm 0.22\pm 0.09)$, and
$0.50\pm0.25^{+0.07}_{-0.04}(0.06\pm0.33\pm0.09)$ for $A_{\phi
K_S}$ and $S_{\phi K_S}$, respectively. These values are
considerably different than the value reported by Belle last year
of \cite{belle03} $S_{\phi K_S} = -0.96\pm 0.50^{+0.09}_{-0.11}$.
 From Figure 3 we see that the current data of $A_{\phi K_S}$ and
$S_{\phi K_S}$ can be easily accommodated by the allowed ranges.
We also note that the allowed ranges can cover last year's Belle
data. Since the error bars on the data are large, no definitive
conclusions can be drawn at present.

\subsection{CP asymmetries in $B\to \phi K_S$ and $B\to \eta' K_S$
in MSSM}

The time dependent CP asymmetry in $B\to \phi K_S$
\begin{eqnarray} S_{\phi K_S}&=&-0.39\pm
0.41,\,\,\,\,\,2002\,\,\, {\rm World-average}\nnb\\S_{\phi
K_S}&=&-0.15\pm 0.33,\,\,\,\,\, 2003\,\,\, {\rm World-average}
\end{eqnarray} is
especially interesting since it deviates greatly from the SM
expectation \be S_{\phi K_S}=\sin (2 \beta (\phi K_S))= \sin (2
\beta (J/\psi K_S)\!) \! + \! O(\lambda^2 \!)
\end{equation}
where $\lambda \simeq 0.2$ appears in Wolfenstein's
parametrization of the CKM matrix. If the error in the average is
inflated by the scale factor, the 2003 World-average will be
$-0.15\pm 0.69$, which deviates from the SM only by $1.3 \sigma$,
i.e., the statistical basis of the effect is weak at present.
Although the impact of these experimental results on the validity
of CKM and SM is currently statistically limited, they have
attracted much interest in searching for new physics and it has
been shown that the deviation can be understood without
contradicting the smallness of the SUSY effect on $B \to J/\psi
K_S$ in the minimal supersymmetric standard model (MSSM)~
\cite{kane,chw}.

Another experimental result is of the time dependent CP asymmetry
$S_{\eta^{\prime} K_S}$ in $B\to \eta^\prime K_S$
\begin{eqnarray}
S_{\eta^\prime K_S}&=& 0.02 \pm 0.34 \pm 0.03 \hspace{5mm}{\rm BaBar}\nonumber\\
&=& 0.43 \pm 0.27 \pm 0.05 \hspace{5mm}{\rm Belle}
\end{eqnarray} which deviates sizably from the SM
expectation. Although both the asymmetries $S_{\phi K}$ and
$S_{\eta^\prime K}$ are smaller than the SM value, $S_{\phi K}$ is
negative and $S_{\eta^\prime K}$ is positive, which implies that
new CP violating physics affects $B\to \phi K_S$ in a dramatic way
but gives $B\to \eta^{\prime} K_S$ a relatively small effect.
Because the quark subprocess $b\to s\bar{s}s$ contributes to both
$B\to \phi K_S$ and $B\to \eta^\prime K_S$ decays one should
simultaneously explain the experimental data in a model with the
same parameters. It has been done in Ref. \cite{kkou} in a model-
independent way in the supersymmetric (SUSY) framework. In Ref.
\cite{kkou} the analysis is carried out using the naive
factorization to calculate hadronic matrix elements of operators
and the neutral Higgs boson (NHB) contributions are not included.
As we have shown in a letter \cite{chw} that both the branching
ratio (Br) and CP asymmetry are significantly dependent of the
$\alpha_s$ corrections of hadronic matrix elements and NHB
contributions are important in MSSM with middle and large
$\tan\beta$ (say, $>$ 8). In this paper \cite{chw1} we study the $B\to \phi
K_S$ and $B\to \ep K_S$ decays in MSSM by calculating hadronic
matrix elements of operators with QCD factorization approach and
including neutral Higgs boson (NHB) contributions. We calculate
the Wilson coefficients of operators including the new operators
which are induced by NHB penguins at LO using the MIA with double
insertions. We calculate the $\alpha_s$ order hadronic matrix
elements of the new operators for $B\rightarrow \phi K_s$ and
$B\rightarrow \eta^\prime K_s$. We analyze constraints from
relevant experimental data among which the new CDF bound of
$B_s\to \mu^+\mu^-$ is used. It is shown that the recent
experimental results on the time-dependent CP asymmetries in $B\to
\phi K_S$ and $B\to \ep K_S$, $S_{\phi K}$ is negative and
$S_{\eta^{\prime} K}$ is positive but smaller than 0.7, which can
not be explained in SM, can be explained in MSSM if there are
flavor non-diagonal squark mass matrix elements of 2nd and 3rd
generations whose size satisfies all relevant constraints from
known experiments ($B\to X_S\gamma, B_s\to \mu^+\mu^-, B\to X_s
\mu^+\mu^-, B\to X_s g, \Delta M_s$, etc.). In particular, we find
that one can explain the experimental results with a flavor
non-diagonal mass insertion of chirality LL or LR when $\alpha_s$
corrections of hadronic matrix elements of operators are included,
in contrast with the claim in the literature. At the same time,
the branching ratios for the two decays can also be in agreement
with experimental measurements.

\subsection{B decays in model III and TC models }

In Ref.\cite{x01} the authors made a systematic study for the new
physics contributions to the charmless two-body hadronic decays of
B and $B_s$ meson induced by the neutral- and charged-Higgs
penguin diagrams in model III: the third type of the
two-Higgs-doublet models. They found that the new physics
enhancements to the branching ratios of those penguin dominated B
meson decay modes can be significant,about $30\%$ to $50\%$, but
the corresponding new physics corrections to CP violating
asymmetries are generally small. Also in model III, the penguin
contributions to the decays of $b \to s g$, $B \to X_s \gamma $ ,
$X_s \gamma\gamma$, $ V\gamma$ ($V=K^*, \rho, \omega$), $l^+ l^-$
and $B^0-\bar{B}^0$ mixing have been studied, for example, in
Ref.\cite{x02}

In the framework of various technicolor models, many B meson decay
modes, such as the radiative decays $B \to X_s \gamma$ and $B \to
X_s \gamma \gamma$, the semileptonic decays $B \to K^{(*)} l^+
l^-$ and $B \to X_s \nu \bar{\nu}$, the charmless two-body
hadronic decays $B \to h_1 h_2$, have been calculated recently
\cite{lugr}. Strong constraints on the parameter space of
technicolor models have been found from these studies.

 \section{Summary}

 In this B physics workshop, a number of interesting topic of B
 physics is discussed. The heavy quark effective theory and heavy quark
 effective field theory are intensively discussed and compared.
 The two approaches for hadronic B decays, PQCD and QCD
 factorization are also discussed and compared. A lot of questions
 are raised. The hadronic wave function including its twist 3
 part are derived. The contribution of new physics especially
 super symmetry are discussed.
 Most scientists and students benefit a lot
  from the lectures and discussions. For a summary, Prof.
 Yue-Liang Wu writes In his transparencies:
\begin{itemize}
\item
  BOTTOM IS BEAUTY ! $\quad b\leftrightarrow \bar{b}$
 \item  BEAUTY FROM TOP ! $\quad b\leftarrow t$
  \item  BEAUTY TO UP ! $\quad \quad b\rightarrow u$
 \item   BEAUTY TO CHARM !  $\quad b\rightarrow c$
 \item   BEAUTY TO STRANGE !  $\quad b\rightarrow s$
 \item  PHYSICS REMAINS AT BOTTOM , LCQCD/HQEFT/QCDF/PQCD IS VERY USEFUL
 \item  MORE NEW PHYSICS BEYOND SM, PHYSICS IS BEAUTY $\&$ CHARM
\end{itemize}

\acknowledgments

This work was supported in part by the National Science Foundation
of China.  The authors thank C.X. Yue and LiaoNing Normal University for hospitality
during the workshop.


\begin{thebibliography}{27}

\bibitem{ylw} Y.L. Wu, Phys. Rev. {\bf D64} (2001) 016001;
More references may be found in: Y.L. Wu, Plenary talk at Intern.
Conf. on Flavor Physics (ICFP 2001), published in {\bf Flavor
physics}, (2002) 217-236,  (World Scientfic Pub. Co.)
hep-ph/0108155.
\bibitem{NA48}J.R. Batley, et.al., NA48 collaboration, Phys. Lett. {\bf B544}
(2002) 97
\bibitem{KTeV} KTeV Collaboration: A. Alavi-Harati, et al, Phys. Rev. {\bf D67} (2003)
012005
\bibitem{kobayashi:1973fv}{M.}~{Kobayashi} {and}
{T.}~{Maskawa},
  {Prog. Theor. Phys.} \textbf{{49}},
 {652} ({1973}).
\bibitem{W0} Y. L. Wu, Mod. Phys. Lett. A {\bf 8}, 819 (1993).
\bibitem{W1} W. Y. Wang, Y. L. Wu and Y. A. Yan, Int. J. Mod. Phys. {\bf A 15}
   (2000) 1817.
\bibitem{W2} Y. A. Yan, Y. L. Wu and W. Y. Wang, Int. J. Mod. Phys. {\bf A 15}
   (2000) 2735.
\bibitem{W3} W. Y. Wang and Y. L. Wu, Int. J. Mod. Phys. {\bf A 16} (2001) 377.
\bibitem{W4} Y. L. Wu and Y. A. Yan, Int. J. Mod. Phys. {\bf A 16} (2001) 285.
\bibitem{W5} W. Y. Wang and Y. L. Wu, Phys. Lett. {\bf B 515} (2001) 57.
\bibitem{W6} W. Y. Wang and Y. L. Wu, Phys. Lett. {\bf B 519} (2001) 219.
\bibitem{W7} W. Y. Wang and Y. L. Wu, M. Zhong, Phys. Rev. {\bf D67} (2003) 014024.
\bibitem{W8} M. Zhong, Y. L. Wu and W. Y. Wang, Int. J. Mod. Phys. {\bf A18} 1959 (2003).
\bibitem{W9} W. Y. Wang, Y. L. Wu and M. Zhong  J. Phys. {\bf G29}
(2003) 1
\bibitem{W10} Y.L. Wu, Y.A. Yan, M. Zhong, Y.B. Zuo and W.Y. Wang,
     Mod. Phys. Lett. A {\bf 18}, 1303 (2003).
\bibitem{W11} W.Y. Wang, Y.L. Wu, Y.A. Yan, M. Zhong and Y.B. Zuo,
Mod.Phys.Lett.A19:1379-1390,2004; hep-ph/0310266.
\bibitem{W12} Y.B Zuo, Y.A Yan, Y.L Wu, W.Y Wang, to be published, hep-ph/0403078
\bibitem{HQET} H. Georgi, Phys. Lett. {\bf B 240} (1990) 447.
\bibitem{HQS} N. Isgur, M. Wise, Phys. Lett. {\bf B 232} (1989) 113;
   {\bf B 237} (1990) 527; {\bf B 206} (1988) 681.
\bibitem{HQL} M. B. Voloshin, M. A. Shifman, Sov. J. Nucl. Phys.
{\bf 45} (1987) 292; {\bf 47} (1988) 199;  E. Eichten, B. Hill,
Phys. Lett. {\bf B 234} (1990) 511; E. Eichten, Nucl. Phys. Proc.
Suppl. {\bf B 4} (1988) 170.
 \bibitem{MR} T. Mannel, W. Roberts, Z. Ryzak, Nucl. Phys. {\bf B 368} (1992) 204.
 \bibitem{WZ1} Y.L. Wu and Y.F. Zhou, Eur. Phys. J .direct C5:014 (2003); hep-ph/021036
 \bibitem{WZ2} Y.L. Wu and Y.F. Zhou, Eur. Phys. J. {\bf C36} 89 (2004); hep-ph/0403252

\bibitem{lb} G.P. Lepage and S.J. Brodsky, Phys.Lett. B{\bf 87}, 359(1979)
Phys.Rev.{\bf D22},2157(1980), {\it ibid.} {\bf 24}, 1808(1981).

\bibitem{bhl} S.J. Brodsky, T. Huang and G.P. Lepage, in {\it
Particles and Fields-2}, Proceedings of the Banff Summer
Institute, Banff, Alberta, 1981, edited by A.Z. Capri and A.N.
Kamal (Plenum, New York, 1983), P143; G.P. Lepage, S.J. Brodskyk
T.Huang, and P.B. Mackenize, {\it ibid.}, p83; T. Huang, {\it in
Proceedings of XXth International Conference on High Energy
Physics}, Madison, Wisconsin, 1980, edited by L.Durand and L.G.
Pondrom, AIP Conf.Proc.No. 69(AIP, New York, 1981), p1000.

\bibitem{sumrule} V.L. Chernyak and A.R.Zhitnitsky, Nucl.Phys.B{\bf
201}, 492 (1982); Phys. Rep .{\bf 112}, 173(1984); Nucl.Phys.B{\bf
246}, 52(1984); T. Huang, X.D. Xiang, and X.N. Wang,
Chin. Phys. Lett.{\bf 2}, 76(1985); Phys. Rev. D{\bf 35}, 1013 (1987).

\bibitem{mh} B.Q. Ma and T. Huang, J.Phys. G{\bf 21}, 765 (1995);
F.G. Cao, J. Cao, T. Huang and B.Q. Ma, Phys. Rev. D{\bf 55},
7107(1997).

\bibitem{hs} T. Huang and Q.X. Shen, Z. Phys. C{\bf 50},
139(1990); T. Huang, B.Q. Ma and Q.X. Shen, Phys. Rev. D{\bf 49},
1490(1994). T Huang, X.G.Wu and X.H.Wu, hep-ph/0404163,to appear in
Phys. Rev. D.

\bibitem{bf} V.M. Braun and I.E. Filyanov, Z. Phys. C{\bf 44},
157(1989).

\bibitem{hwz} T. Huang and X.H. Wu and M.Z. Zhou,hep-ph/0402100,
Phys. Rev. D{\bf 70},014013;

\bibitem{hw} T. Huang and X.G. Wu, hep-ph/0408252.

\bibitem{Braun2} V. M. Braun and I. B. Filyanov,
    Z. Phys. {\bf C48}, (1990)239.
\bibitem{Ball} P. Ball, V. M. Braun, Y. Koike and K. Tanaka,
    Nucl. Phys. {\bf B529}, (1998)323; hep-ph/9802299.

\bibitem{KLS} Y.Y. Keum, H-n. Li, and A.I. Sanda,
Phys. Lett. B {\bf 504}, 6 (2001); Phys. Rev. D {\bf 63}, 054008
(2001); Y.Y. Keum and H-n. Li, Phys. Rev. {\bf D63}, 074006
(2001).
\bibitem{beneke} M.~Beneke, G.~Buchalla, M.~Neubert and C.~T.~Sachrajda,
Nucl.\ Phys.\ B {\bf 591}, 313 (2000).

\bibitem{grozin} A.~G.~Grozin and M.~Neubert,
Phys.\ Rev.\ D {\bf 55}, 272 (1997).


\bibitem {qiao}H. Kawamura, J. Kodaira,
C.-F. Qiao., K. Tanaka Phys.\ Lett.\ {\bf{B} 523} (2001) 111; MPLA
18 (2003) 799

\bibitem{cdl} C. D. L\"u, Eur. Phys. J. C{24} (2002) 121; Z.T. Wei
and M.Z. Yang, Nucl. Phys. B642 (220) 263.
\bibitem{BSW} M. Bauer, B. Stech, M. Wirbel,
Z. Phys. C {\bf 29}, 637 (1985); {\sl ibid.} {\bf 34}, 103 (1987).

\bibitem{cz}V.L. Chernyak and I.R. Zhitnitsky,
                        Nucl. Phys. {\bf B345}, 137(1990).
\bibitem{isg} N. Isgur and C.H Llewelyn-Smith, Phys. Rev. Lett.{\bf 52},1080 (1984);
Nucl. Phys. {\bf B317}, 526(1989).
\bibitem{rad} A. V. Radyushkin, Acta Phys. Pol.15, 403(1984).
\bibitem{stef} N. G. Stefanis, hep-ph/9911375.
\bibitem{BBNS} M. Beneke, G. Buchalla, M. Neubert, and C.T. Schrajda,
               Phys. Rev. Lett. {\bf 83}, 1914 (1999).
\bibitem{BSS}M. Bander, D. Silverman, and A. Soni, Phys. Rev. Lett, {\bf 43},
        242 (1979).

\bibitem{pdg} Particle Data Group, Phys. Lett. {\bf B592} (2004) 1.

\bibitem{ch} J. Chay and C. Kim, hep-ph/0009244;
 H.Y.~Cheng and K.C.~Yang, Phys.\ Rev.\ D {\bf 63} (2001)074011.

\bibitem{song1} Z. Song, C. Meng and K.T. Chao, Eur. Phys. J. C36 (2004) 365.

\bibitem{song2} Z. Song and K.T. Chao, Phys. Lett. B568 (2003)127.

\bibitem{song3} Z. Song, C. Meng, Y.J.Gao and K.T. Chao,
Phys. Rev. D {\bf 69} (2004)054009.

\bibitem{gao} Y.J. Gao, C. Meng, and K.T. Chao, to be submitted.

\bibitem{bbl}  G.T. Bodwin, L. Braaten, and G. P. Lepage, Phys. Rev. D51, 1125 (1995).
\bibitem{mix} Y.P.Kuang and T.M.Yan, Phys. Rev. D41, 155 (1990);

 Y.B.Ding, D.H.Qin, and K.T.Chao, Phys. Rev. D44, 3562 (1991);

J.L.Rosner, Phys. Rev. D64, 094002 (2001);  K.Y. Liu and K.T.
Chao, hep-ph/0405126.

\bibitem{belled} Belle Collaboration,  R. Chistov et al.,
Phys. Rev. Lett. 93 (2004) 051803.
\bibitem{yuanko}F.Yuan, C.F.Qiao, and K.T.Chao, Phys. Rev. D56, 329
(1997); P.W.Ko, J.Lee, and H.S.Song, Phys. Lett. B395, 107(1997).


\bibitem{BS} J. Botts and G. Sterman, Nucl. Phys. {\bf B225}, 62 (1989).
\bibitem{LS} H-n. Li and G. Sterman, Nucl. Phys. {\bf B381}, 129
(1992).
\bibitem{LY1} H-n. Li and H.L. Yu, Phys. Rev. Lett. {\bf 74}, 4388 (1995);
Phys. Lett. B {\bf 353}, 301 (1995); Phys. Rev. D {\bf 53}, 2480
(1996).
\bibitem{CL} C.H. Chang and H-n. Li, Phys. Rev. D {\bf 55}, 5577 (1997).
\bibitem{YL} T.W. Yeh and H-n. Li, Phys. Rev. D {\bf 56}, 1615 (1997).
\bibitem{CLY} H.Y. Cheng, H-n. Li, and K.C. Yang,
Phys. Rev. D {\bf 60}, 094005 (1999).
\bibitem{NL} H-n. Li, Phys. Rev. D {\bf 64}, 014019 (2001); M. Nagashima
and H-n. Li, hep-ph/0202127; Phys. Rev. D {\bf 67}, 034001 (2003).

\bibitem{LUY} C. D. L\"{u}, K. Ukai, and M. Z. Yang,
Phys. Rev. D {\bf 63}, 074009 (2001); C. D. L\"{u} and M. Z. Yang,
Eur. Phys. J. C23, 275 (2002).
\bibitem{KS} Y.Y. Keum and A. I. Sanda, Phys. Rev. D {\bf 67}, 054009
(2003).
\bibitem{Keum02} Y.Y. Keum, hep-ph/0210127.
\bibitem{Keum01} Y.Y. Keum, hep-ph/0209208.
\bibitem{NPe} M. Neubert and A.A. Petrov, Phys. Lett. B {\bf 519}, 50
(2001).
\bibitem{TLS2} T. Kurimoto, H-n. Li, and A.I. Sanda,
Phys. Rev. D {\bf 67}, 054028 (2003).
\bibitem{SHB} A.P. Szczepaniak, E.M. Henley, and S.J. Brodsky,
Phys. Lett. B {\bf 243}, 287 (1990); G. Burdman and J.F. Donoghue,
Phys. Lett. B {\bf 270}, 55 (1991).

\bibitem{TLS} T. Kurimoto, H-n. Li, and A.I. Sanda,
Phys. Rev. D {\bf 65}, 014007, (2002).
\bibitem{BelleC} Belle Colla., K. Abe {\it et al.},
Phys. Rev Lett. {\bf 88}, 052002 (2002).
\bibitem{CLEOC} CLEO Colla., T.E. Coan {\it et al.},
Phys. Rev. Lett. {\bf 88}, 062001 (2002). Phys. Rev. D {\bf 59},
092004 (1999).
\bibitem{Bab} BaBar Colla., B. Aubert  {\it et al.}, hep-ex/0207092.
\bibitem{GKKP} M. Gourdin, A.N. Kamal, Y.Y. Keum and X.Y. Pham,
Phys. Letts. {\bf B 333}, 507 (1994).

\bibitem{deta} C.D. L\"u,
Phys. Rev. D68, 097502 (2003).

\bibitem{0304132} M. Battaglia {\it et al.}, {\em The CKM
Matrix and the Unitarity Triangle}, hep-ph/0304132, and references
therein.

\bibitem{0406184}
J.Charles {\it et al.}, {\em CP Violation and the CKM Matrix:
Assessing the Impact of the Asymmetric  B Factories},
hep-ph/0406184.

\bibitem{hfag04}
Heavy Flavor Averaging Group,
http://www.slac.stanford.edu/xorg/hfag/; 2004.

\bibitem{0406082ex}
M.Verderi, hep-ex/0406282, BaBar-Talk-04/011.

\bibitem{phiks}
Y.L.Wu and Y.F.Zhou, {\bf Eur.Phys.J. C}36 (2004)89; J.F.~Cheng,
C.S.~Huang, and X.H.~Wu, {\bf Phys.Lett. B}585, 287 (2004).

\bibitem{alpha}
C.D.L\"u and Z.J.Xiao, {\bf Phys.Rev.D}66 (2002)074011; Z.J.Xiao,
C.D.L\"u and L. Guo, hep-ph/0303070 and references therein.

\bibitem{g04}
M.Gronau and J.L. Rosner, {\bf Phys. Lett. B}595 (2004)339; For a
brief reveiw, see M.Gronau, hep-ph/0407316.

\bibitem{gamma}
R.Fleischer, {\bf Phys. Rep.} 370 (2003)257; hep-ph/0405091;

\bibitem{xiao02}
Z.J.Xiao and M.P.Zhang, {\bf Phys.Rev. D}65 (2002)114017.

\bibitem{pi04}
A.J.Buras {rt al.,} {\bf Phys.Rev.Lett.} 92 (2004)101804;
G.Buchalla and A.S.Safir, {\bf Phys.Rev.Lett.} 93 (2004)021801.

\bibitem{BS-in} Chao-Hsi Chang, Jiao-Kai Chen,
Xue-Qian Li and Guo-Li Wang, hep-ph/0406050.

\bibitem{brei-e} Chao-Hsi Chang and Jiao-Kai Chen, {\it The Instantaneous
Bethe-Salpeter Equation and Its Analog: the Breit-like Equation}
in preparation.

\bibitem{expcpasy}
 BABAR COLLABORATION, B. Aubert et al.,
hep-ex/0407057; Belle Collaboration, Y. Chao et al.,
hep-ex/{0408100}.

\bibitem{hfag}
Heavy Flavor Averaging Group [HFAG],
http://www.slac.stanford.edu/xorg/hfag

\bibitem{cpgamma}
BABAR COLLABORATION, B. Aubert et al., PHYS. REV. LETT.
{93}{2004}{021804}; Belle Collaboration, S. Nishida et al., PHYS.
REV. LETT. {93}{2004}{031803}.

\bibitem{babar} M. Giorgi, Plenary talk give at ICHEP'04;
Y. Sakai, Plenary talk given at ICHEP'04.

\bibitem{hly} X.-G. He, C.-S. Li and L.-L. Yang, arXiv: hep-ph/0409338.


\bibitem{sm} G. Buchalla, A. Buras and M. Lautenbacher, Rev. Mod.
Phys. {\bf 68} (1996) 1125; A. Buras, M. Jamin and M.
Lautenbacher, Nucl. Phys. {\bf B400} (1993) 75; M. Ciuchini et
al., ibid. {\bf B415}, 403(1994); N. Deshpande and X.-G. He, Phys.
Lett. {\bf B336} (1994)471.

\bibitem{sm1}  K. Chetyrkin, M. Misiak and M. M\" unz,
Phys. Lett. {\bf B400} (1997) 206.

\bibitem{kagan}  A.L. Kagan and M. Neubert,
Phys. Rev. D{\bf 58} (1998) 094012.


\bibitem{susydipole}  S. Bertolini, F. Borzumati and A. Masiero,
Nucl. Phys. {\bf B284} (1987) 321; S. Bertolini {\it et al.}, {\it
ibid.} {\bf B353} (1991) 591.


\bibitem{massinsertion}  F. Gabbiani et al., Nucl. Phys. {\bf
477} (1996) 321; A. Kagan, Phys. Rev. {\bf D51} (1995) 6196.

\bibitem{pheno}
C.-K. Chua, X.-G. He and W.-S. Hou, Phys. Rev. {\bf D60},
014003(1999); X.-G. He, J.-Y. Leou and J.-Q. Shi, Phys. Rev. {\bf
D64} (2001) 094018.

\bibitem{bbns} M. Beneke and M. Neubert, Nucl. Phys. {\bf B675}
(2003) 333; M. Beneke, G. Buchalla, M. Neubert, C. T. Sachrajda,
Nucl. Phys. {\bf B606} (2001) 245.

\bibitem{atwood}  D. Atwood, M. Gronau and A. Soni,
Phys. Rev. Lett. {\bf 79} (1997) 185.

\bibitem{phiK} P. Ko et al., Phys. Rev. Lett. {\bf 90} (2003) 141803;
C.-K. Chua, W.-S. Hou,  and M. Nagashima, Phys. Rev. Lett. {\bf
92} ( 2004)201803.


\bibitem{kgamma}
BABAR COLLABORATION, B. Aubert et al.,  hep-ex/{0405082}.

\bibitem{kphi}
BABAR COLLABORATION, B. Aubert et al.,  hep-ex/{0408072}; Belle
Collaboration, K, Abe et al.,
 hep-ex/{0409049}.


\bibitem{ss} BABAR COLLABORATION, B. Aubert et al.,  arXiv: hep-ex/0407003;
Belle Collaboration M. Nakao et al.,  arXiv: hep-ex/0402042.

\bibitem{belle03} Belle Collaboration, K. Abe et al.,  Phys. Rev. Lett. {\bf 91},
(2003)261602.

\bibitem{kane}G.L. Kane et al., hep-ph/0212092,
 Phys. Rev. Lett. {\bf 90} (2003) 141803.
\bibitem{chw}J.-F. Cheng, C.-S. Huang and X.-H. Wu, Phys. Lett. {\bf B 585}
 (2004) 287 [arXiv:hep-ph/0306086].

\bibitem{kkou}S.~Khalil, E.~Kou, Phys. Rev. Lett. {\bf 91} (2003) 241602.
\bibitem{chw1}J.-F. Cheng, C.-S. Huang and X.-H. Wu,
hep-ph/0404055, to appear in Nucl. Phys. B.
\bibitem{x01}
Z.J.~Xiao, C.S.~Li and K.T.~Chao, Phys. ReV. D 63, 074005(2001);
Phys. ReV. D 65, 114021 (2002) and references therein; D.~Zhang,
Z.J.~Xiao and C.S.~Li, Phys. ReV. D 64, 014014(2001).

\bibitem{x02}
Z.J.~Xiao, C.S.~Li and K.T.~Chao, Phys. Lett. B 473, 148 (2000);
Phys. ReV. D 62, 094008(2000); J.J.~Cao, Z.J.~Xiao and G.R.~Lu,
Phys. ReV. D 64, 014012(2001); C.S.~Huang and S.H.~Zhu, Phys. ReV.
D 68, 114020 (2003); Y.B.~Dai, C.S.~Huang, J.T.~Li and W.J.~Li,
Phys. ReV. D 67, 096007 (2003); Z.J.~Xiao and C.~Zhuang, Eur.
Phys. J. C 33, 349(2004); Z.J.~Xiao and Libo~Guo, Phys. ReV. D 69,
014002 (2004).

\bibitem{lugr}
G.R.~Lu {\it et al.}, Phys. ReV. D 54, 5647 (1996); G.R.~Lu,
Z.H.~Xiong and Y.G.~Cao, Nucl. Phys.  B 487, 43(1997); G.R.~Lu,
Z.J.~Xiao, H.K.~Guo and L.X.~L\"u, J. Phys.  G25 (1999)L85;
Z.J.~Xiao, W.J.~Li, Libo Guo and G.R.~Lu, Eur. Phys. J. C 18, 681
(2001); Z.H.~Xiong, J.M.~Yang, Phys. Lett. B 546, 221 (2002);
Z.J.~Xiao, C.D.~L\"u and W.J.~Huo, Phys. ReV. D 67, 094021 (2003).

\end{thebibliography}
\end{document}